\newif\ifonecol
\onecoltrue

\ifonecol
    \documentclass[12pt,draftclsnofoot,onecolumn]{IEEEtran}
\else
    \documentclass[journal,comsoc]{IEEEtran}
\fi

\usepackage{amsmath,graphics,amssymb,epsfig,subfigure,color,amsthm,cite}
\usepackage{array}
\usepackage{multirow}
\usepackage{enumerate}
\usepackage{algorithmic}
\usepackage{algorithm}
\usepackage{bm}
\usepackage{float}

\newtheorem{prop}{Proposition}

\newcommand{\argmin}{\operatornamewithlimits{argmin}}
\makeatletter
\newcommand{\vast}{\bBigg@{4.5}}
\newcommand{\Vast}{\bBigg@{7.5}}
\makeatother

\begin{document}
\ifonecol
	\title{\LARGE{Soft-Output Detection Methods for Sparse Millimeter Wave MIMO Systems with Low-Precision ADCs}}
\else
	\title{Soft-Output Detection Methods for Sparse  Millimeter Wave MIMO Systems with Low-Precision ADCs}
\fi

\author{Yo-Seb Jeon, Heedong Do, Song-Nam Hong, and Namyoon Lee
	\thanks{Y.-S. Jeon, H. Do, and N. Lee are with the Department of Electrical Engineering, POSTECH, Pohang, Gyeongbuk 37673, South Korea  (e-mail: \{yoseb.jeon,doheedong,nylee\}@postech.ac.kr).}
	\thanks{S.-N. Hong is with the Department of Electrical and Computer Engineering, Ajou University, Suwon, Gyeonggi 16499, South Korea (e-mail: snhong@ajou.ac.kr).}
	%\thanks{This work was supported in part by Samsung Research Funding $\&$ Incubation Center of Samsung Electronics under Project Number SRFC-IT1702-00, and in part by the National Science Foundation under Grant No. NSF-CCF-1527079. This work was presented in part at the 2018 IEEE 87th Vehicular Techonology Conference (VTC2018-Spring).}
}
\vspace{-2mm}	

\maketitle
\vspace{-12mm}

\begin{abstract} % up to 200 words
The use of low-precision analog-to-digital converters (ADCs) is a low-cost and power-efficient solution for a millimeter wave (mmWave) multiple-input multiple-output (MIMO) system operating at sampling rates higher than a few Gsample/sec. This solution, however, can make significant frame-error-rates (FERs) degradation due to inter-subcarrier interference when applying conventional frequency-domain equalization techniques. In this paper, we propose computationally-efficient yet near-optimal soft-output detection methods for the coded mmWave MIMO systems with low-precision ADCs. The underlying idea of the proposed methods is to construct an extremely sparse inter-symbol-interference (ISI) channel model by jointly exploiting the delay-domain sparsity in mmWave channels and a high quantization noise caused by low-precision ADCs. Then we harness this sparse channel model to create a trellis diagram with a reduced number of states or a factor graph with very sparse edge connections. Using the reduced trellis diagram, we present a soft-output detection method that computes the log-likelihood ratios (LLRs) of coded bits by optimally combining the quantized received signals obtained from multiple receive antennas using a forward-and-backward algorithm. To reduce the computational complexity further, we also present a low-complexity detection method using the sparse factor graph to compute the LLRs in an iterative fashion based on a belief propagation algorithm. Simulations results demonstrate that the proposed soft-output detection methods provide significant FER gains compared to the existing frequency-domain equalization techniques in a coded mmWave MIMO system using one- or two-bit ADCs.
\end{abstract}


\section{Introduction}\label{sec1}
%%%%%% Motivation of using low-precision ADC in mmillimeter wave
MmWave communication combined with massive multiple-input multiple-output (MIMO) is a key feature of next-generation wireless systems to provide high data rates beyond hundreds of Gbits/sec \cite{Swindle:14,Sun:12,Han:15}. Thanks to relatively large bandwidths available at the mmWave band, it is possible to linearly increase the throughput of the wireless system with the bandwidth. In addition, the use of a massive antenna array  allows the system to compensate a significant path loss at the mmWave band by beamforming gains. In spite of the significant rate enhancement, implementing the mmWave system that uses both the large bandwidth and the massive antenna array is difficult. One of the major reasons is that prohibitive power consumption is required by high-precision (8$\sim$16 bits) analog-to-digital converters (ADCs) at the receiver, whose power consumption increases linearly with both the system bandwidth (i.e., the sampling rate) and the number of RF chains \cite{Murmann,Walden:99,Singh:09}. A simple yet effective solution to resolve this difficulty is to reduce the number of precision bits of the ADCs \cite{Mezghani:07,Madhow:09,Bjornson:15,Zhang:16,Zhang:17,Zhang:18,Mo:18}, because the power consumption of the ADCs decreases  exponentially with the number of quantization bits \cite{Murmann,Walden:99}.
%the prohibitive power consumption required by high-precision (8$\sim$16 bits) analog-to-digital converters (ADCs) , as its power consumption increases linearly with both the system bandwidth (i.e., the sampling rate) and the number of RF chains \cite{Murmann,Walden:99,Singh:09}. A simple yet effective solution to resolve this difficulty is to reduce the number of precision bits of the ADCs \cite{Mezghani:07,Madhow:09,Bjornson:15,Zhang:16,Zhang:17,Zhang:18,Mo:18}, because the power consumption of the ADCs decreases  exponentially with the number of quantization bits \cite{Murmann,Walden:99}.}

Unfortunately, the use of low-precision (1$\sim$2 bits) ADCs faces a challenge brought by the nonlinear quantization effect of the ADCs. Particularly, in a coded system, this nonlinear effect causes a severe frame-error-rate (FER) degradation due to inter-subcarrier interference when applying conventional frequency-domain equalization techniques such as orthogonal frequency division modulation (OFDM) or single-carrier frequency domain equalization (SC-FDE). To resolve this problem, it is essential to design effective soft-output detection methods for mmWave (frequency-selective) MIMO systems when low-precision ADCs are employed. In this paper, we make progress toward designing near-optimal time-domain soft-output detection methods.  Using the sparse property in the mmWave channels \cite{Mo:18,Akdeniz:14,Rappaport:15,Samimi:16}, we present how to extract out the soft-information (e.g., log-likelihood ratios (LLRs) of coded bits) by optimally combining the quantized received signals obtained from multiple receive antennas in a computationally-efficient manner.

\subsection{Related Work}
%%%%%%% Detections: Frequency-flat -> Frequency-selective -> Soft-output (Studer and S.-N. Hong) -> Optimal soft-output?
There is a rich literature on data detection methods in MIMO systems with low-precision ADCs \cite{Wang:14,Choi:16,Hong:18,Jeon:TWC:18,Wen:16,Jeon:TVT:18,Jeon:WCNC:18,Hyowon:18}. For frequency-flat MIMO channels, the maximum-likelihood (ML) detection and its low-complexity variations were introduced in \cite{Wang:14,Choi:16,Hong:18,Jeon:TWC:18}. Data detection methods that are robust to the effect of a high channel estimation error were also proposed using several approaches such as Bayesian approach for joint channel-and-data estimation \cite{Wen:16}, supervised-learning approach \cite{Jeon:TVT:18}, and reinforcement-learning approach \cite{Jeon:WCNC:18}. Unfortunately, these methods are not applicable to general mmWave MIMO channels with frequency-selectivity because of the frequency-flat assumption that ignores the effect of inter-symbol interference (ISI). Recently, a data detection method for frequency-selective channels was proposed based on Viterbi algorithm \cite{Hyowon:18}. This method is shown to be optimal in the sense of detecting the sequence of transmitted data symbols. The common limitation of the aforementioned detection methods is that they cannot produce the LLRs of coded bits, which are the necessary inputs for modern channel decoders (e.g., Turbo, low-density-parity-check (LDPC) and polar codes) to obtain the optimal coding gain.

Soft-output detection methods for conventional MIMO systems with high-precision ADCs have been intensively studied in the literature \cite{BCJR:74,Li:95,Kurkoski:02,Colavolpe:05,Kaynak:05,Zhang:15}. Frequency-domain equalization techniques with a soft demapper were popular, because they allow the computation of LLRs with per-subcarrier operation. Whereas, the time-domain soft-output detection methods were not preferable for the conventional MIMO systems due to their high computational complexity. For example, the BCJR algorithm using the forward-backward recursion in \cite{BCJR:74,Li:95}  computes the exact LLRs based on the Trellis diagram constructed by a ISI channel. This algorithm requires the computational complexity that increases exponentially with the number of ISI channel taps, the modulation size, and the number of transmit antennas. To reduce the complexity, soft-output detection methods based on the belief propagation (BP) algorithm were also proposed in \cite{Kurkoski:02,Colavolpe:05,Kaynak:05}. They compute the the LLR values using an iterative message-passing algorithm based on the factor graph constructed by a ISI channel. Unfortunately, both algorithms cannot be directly applicable to mmWave MIMO systems with low-precision ADCs. The major challenge is that in these systems, only quantized observations of the received signals are available at the detector to compute the LLRs, which are distorted by the nonlinear quantization effect.

Very limited work has focused on the development of soft-output detection methods for MIMO systems with low-precision ADCs \cite{Hong:CL:18,Studer:16,Cao:17,Wang:17,He:18,Sun:18}. In our prior work \cite{Hong:CL:18}, a weighted Hamming distance was used to compute the LLRs for the MIMO systems with one-bit ADCs; yet, this work does not take into account the ISI effect of mmWave channels. Soft-output detection methods for frequency-selective channels were proposed in \cite{Studer:16,Cao:17,Wang:17,He:18} based on the frequency-domain equalization (e.g., OFDM and SC-FDE). Unlike conventional OFDM/SC-FDE systems, per-subcarrier soft-output detection is highly suboptimal in mmWave OFDM/SC-FDE systems with low-precision ADCs. The major reason is that when the fast Fourier transform operation is applied after the ADCs, perfect inter-subcarrier interference cancellation is not feasible due to the nonlinearity of the quantization function. To resolve this problem, a joint-subcarrier detection method based on convex optimization was developed in \cite{Studer:16}, while iterative detection algorithms based on approximations were considered in \cite{Cao:17,Wang:17,He:18}. These frequency-domain techniques were shown to be fairly effective when the number of receive antennas is sufficiently larger than the number of simultaneously transmitted data streams at the transmitter. Recently, a joint soft-output detection and channel-decoding method has been developed in \cite{Sun:18} on the basis of bilinear GAMP algorithm, but the algorithm is limited to the use for single-input single-output (SISO) systems. Moreover, none of the aforementioned methods in \cite{Studer:16,Cao:17,Wang:17,He:18,Sun:18} guarantees the optimality in the soft-output detection performance, because all these methods compute the LLRs based on the approximate algorithms.

\subsection{Contributions}
The major contributions of this paper are summarized as follows:

\begin{itemize}
    \item We construct an extremely sparse ISI channel model for mmWave MIMO systems with low-precision ADCs, by jointly exploiting the delay-domain sparsity in mmWave channels and a high quantization noise caused by low-precision ADCs. Considering the quantization noise level, the constructed channel model consists only of a few \textit{dominant} channel-impulse-response (CIR) taps, while treating weak CIR taps as additional noise. We also develop a dominant-tap-selection algorithm to reduce a modeling error in the constructed channel. The key idea of the developed algorithm is to minimize the normalized mean-squared-error between the arguments of two conditional probability mass functions (PMFs), computed based on the true channel model and on the extremely sparse channel model, respectively. The design parameters of the developed algorithm are chosen to adjust the performance-complexity tradeoff of the soft-output detection. 
    %One attractive feature is that the criterion of the algorithm is a function of the CIR-tap power, the noise level, and the quantization function at the ADCs.  
	
    \item We propose a soft-output detection method, referred to as \textit{quantized BCJR (Q-BCJR)}, that computes the LLRs of coded bits by optimally combining the quantized received signals obtained from multiple receive antennas using the forward-and-backward algorithm. Based on the extremely sparse ISI channel model, we reduce the computational complexity of Q-BCJR by creating a trellis diagram that has a reduced number of states determined only by the dominant CIR taps. From the complexity analysis, we show that the computational complexity order of Q-BCJR depends only on the maximum delay index of the dominant CIR taps. One promising feature of Q-BCJR is that it guarantees near-optimal performance when the power of the weak CIR taps is sufficiently lower than the noise level. In addition, for the extreme case (i.e., every CIR tap is dominant), Q-BCJR becomes the optimal soft-output detection method that computes the exact LLRs at the expense of the computational complexity. 
    
    \item We also propose a low-complexity soft-output detection method, referred to as \textit{quantized belief propagation (Q-BP)}, that iteratively compute the LLRs using the BP algorithm. Based on the extremely sparse ISI channel model, we reduce the computational complexity of Q-BP by constructing a sparse factor graph that ignores the edges associating with the weak CIR taps. We also design the messages of Q-BP that consider not only the quantization function at the ADCs but also the effect of the ignored edges. From the complexity analysis, we show that the computational complexity order of Q-BP depends only on the number of the dominant CIR taps, which achieves a significant reduction in the computational complexity compared to Q-BCJR. 
    
    \item Using simulations, we evaluate the frame-error-rate (FER) performance of the proposed soft-output detection methods for a coded mmWave MIMO system with low-precision ADCs, compared to the existing OFDM-based detection methods. Simulation results show that both Q-BCJR and Q-BP outperform the existing methods in terms of FERs when employing one- or two-bit ADCs. It is also shown that both the proposed methods are robust to a channel estimation error when applying a practical channel-estimation method. By simulations, we also show that our dominant-tap-selection algorithm effectively improves the performance-complexity tradeoff achieved by the proposed methods.  
\end{itemize}

\subsubsection*{Notation}
Upper-case and lower-case boldface letters denote matrices and column vectors, respectively.
$\mathbb{E}[\cdot]$ is the statistical expectation,
$\mathbb{P}(\cdot)$ is the probability,
$(\cdot)^\top$ is the transpose,
$(\cdot)^{\sf H}$ is the conjugate transpose,
$|\cdot|$ is the absolute value,
${\sf Re}\{\cdot\}$ is the real part,
${\sf Im}\{\cdot\}$ is the imaginary part,
$\|{\bf a}\|\!=\!\sqrt{{\bf a}{\bf a}^{\sf H}}$ is the Euclidean norm of a vector ${\bf a}$, 
and $\|{\bf A}\|_{\rm F}\!=\!\sqrt{{\sf Tr}({\bf A}{\bf A}^{\sf H})}$ is the Frobenius norm of a matrix ${\bf A}$.
$\mathbb{I}\{\mathcal{A}\}$ is an indicator function which equals one if an event $\mathcal{A}$ is true and zero otherwise.
${\bf 0}_n$ is an $n$-dimensional vector whose elements are zero.	
$\Phi(\cdot)$ is the cumulative distribution of the standard normal random variable.

\section{System Model and Preliminary}
\begin{figure*}
	\centering \vspace{-0.3cm}
	\includegraphics[width=16cm]{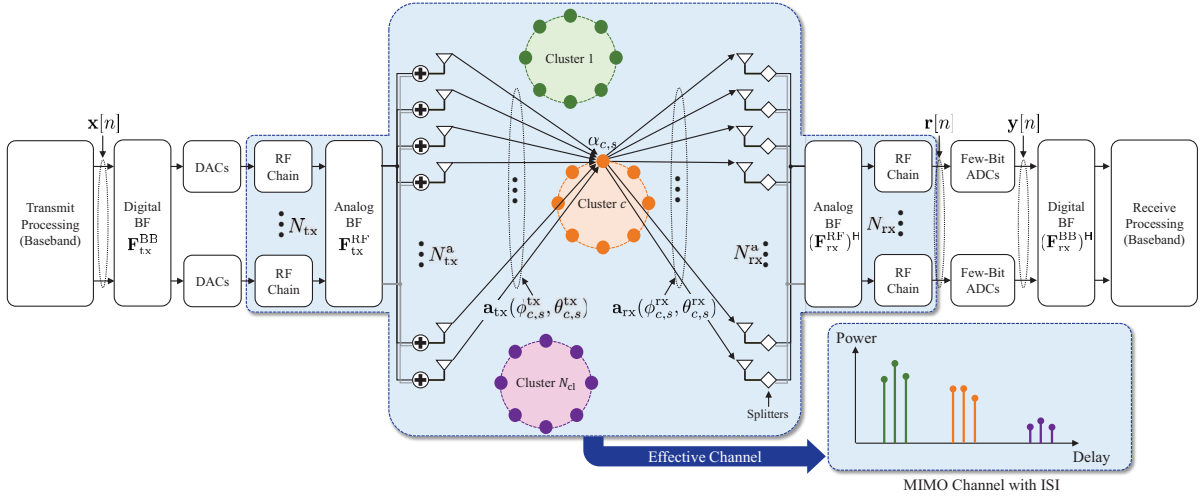} \vspace{-0.1cm}\caption{A mmWave MIMO communication system with low-precision ADCs when hybrid beamforming is employed at both a transmitter and a receiver.} \label{fig:System}
\end{figure*}

We consider a mmWave MIMO communication system with low-precision ADCs, as illustrated in Fig.~\ref{fig:System}. In the considered system, a transmitter equipped with ${N}_{\rm tx}^{\rm a}$ antenna elements followed by  $N_{\rm tx} \leq {N}_{\rm tx}^{\rm a}$ RF chains communicates with a receiver equipped with ${N}_{\rm rx}^{\rm a}$ antenna elements followed by $N_{\rm rx} \leq {N}_{\rm rx}^{\rm a}$ RF chains.

%%%%%%%%%%%%%%%%%%%%%%%%%%%%%%%%%%%%%%%%%%%%%%%%%%%%%%%%%%%%%%%%%%%%%%%%%%%%%%%%%%%%%%%%%%%%
%%%%%%%%%%%%%%%%%%%%%%%%%%%%%%%%%%%%%%%%%%%%%%%%%%%%%%%%%%%%%%%%%%%%%%%%%%%%%%%%%%%%%%%%%%%%
\subsection{Channel Model}
A mmWave channel between the transmitter and the receiver is modeled using a transmit array-response vector, a receive array-response vector, and $N_{\rm cl}$ multi-path clusters, in which the $c$-th cluster consists of $N_{{\rm path},c}$ subpaths. Let ${\bf a}_{\rm tx}(\phi,\theta) \in \mathbb{C}^{{N}_{\rm tx}^{\rm a}}$ and ${\bf a}_{\rm rx}(\phi,\theta)\in \mathbb{C}^{{N}_{\rm rx}^{\rm a}}$ be a transmit and a receive array-response vector, respectively, which depends on the geometry of the antenna elements, a horizontal angle $\phi$ of arrival (or departure), and a vertical angle $\theta$ of arrival (or departure). Let also $\alpha_{c,s}\in \mathbb{C}$ and $\tau_{c,s}\in\mathbb{R}$ be the complex channel gain and the propagation delay of the $s$-th subpath in the $c$-th cluster, respectively. Then an analog channel matrix at discrete time index $\ell$, namely ${\bf A}[\ell]\in\mathbb{C}^{{N}_{\rm rx}^{\rm a}\times{N}_{\rm tx}^{\rm a}}$, is expressed as
\begin{align}\label{eq:mmWave_ch}
	{\bf A}[\ell] = \sum_{c=1}^{N_{\rm cl}}\sum_{s=1}^{N_{{\rm path},c}}\alpha_{c,s} {\bf a}_{\rm rx}(\phi_{c,s}^{\rm rx},\theta_{c,s}^{\rm rx})
	{\bf a}_{\rm tx}^{\sf H}(\phi_{c,s}^{\rm tx},\theta_{c,s}^{\rm tx})p(\ell T_{\rm s} \!-\! \tau_{c,s}),~\ell \in\{0,\ldots,L-1\},
\end{align}
where  $T_{\rm s}$ is the symbol duration, $p(\cdot)$ is a pulse-shaping function, $\phi_{c,s}^{\rm tx}$ ($\theta_{c,s}^{\rm tx}$) is a horizontal (vertical) angle of departure, $\phi_{c,s}^{\rm rx}$ ($\theta_{c,s}^{\rm rx}$) is a horizontal (vertical) angle of arrival associating with the $s$-th subpath in the $c$-th cluster, $L=\lfloor \frac{ \tau_{\rm max}}{T_{\rm s}}+\frac{1}{2} \rfloor$ is the maximum delay index of the analog channel, and $\tau_{\rm max}=\max_{c,s} \tau_{c,s}$ is the maximum propagation delay.

\begin{figure*}
	\centering 
	\subfigure[CDF of $L$ and $|\mathcal{S}|$]
	{\epsfig{file=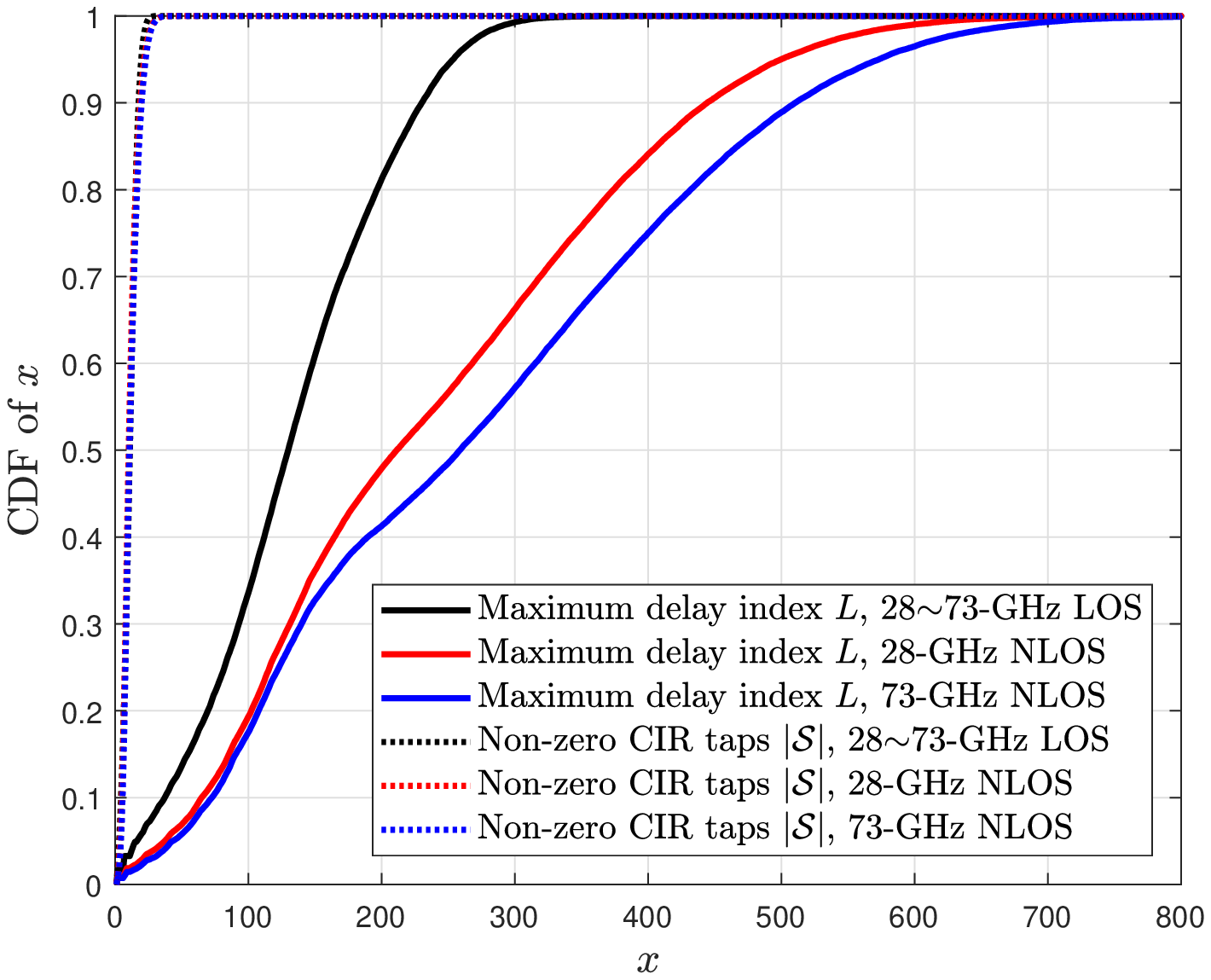, width=7cm}}
	\subfigure[A typical power-delay distribution of a 28-GHz NLOS channel]
	{\epsfig{file=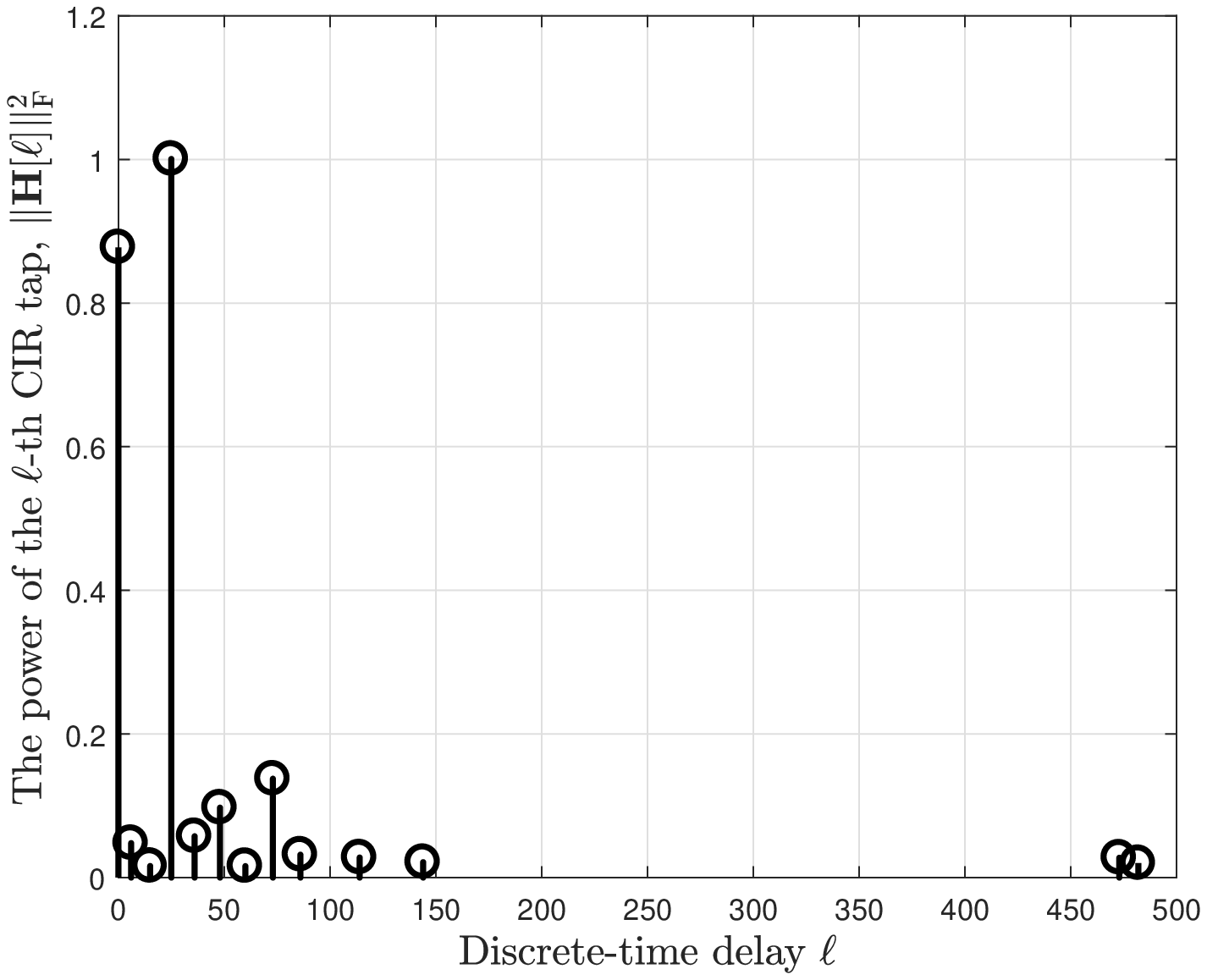, width=7cm}}
	\caption{(a) The cumulative distribution function (CDF) of $L$ and $|\mathcal{S}|$ for various mmWave channels implemented accroding to \cite{Samimi:16}, and (b) a typical power-delay distribution of a 28-GHz NLOS channel.} \vspace{-3mm}
	\label{fig:TapDist}
\end{figure*}

We consider an effective mmWave channel that contains both the transmit and receive analog BFs, as illustrated in Fig.~\ref{fig:System}. %This channel can be modeled by an $L$-tap channel-impulse-response (CIR), where $L$ is the number of CIR taps determined by the ratio between the maximum delay spread $\tau_{\rm max}=\max_{c,s} \tau_{c,s}$ and the symbol duration $T_{\rm s}$, i.e., $L=\lfloor \frac{ \tau_{\rm max}}{T_{\rm s}}+\frac{1}{2} \rfloor$. 
Let  ${\bf F}_{\rm tx}^{\rm RF}\in\mathbb{C}^{{N}_{\rm tx}^{\rm a}\times N_{\rm tx}}$ and ${\bf F}_{\rm rx}^{\rm RF}\in\mathbb{C}^{{N}_{\rm rx}^{\rm a}\times N_{\rm rx}}$ be the analog BF matrix at the transmitter and the receiver, respectively, that consists of phase shifters. 
	%Since we focus on the design of the soft-output detector for the given channel realization, we simply adopt the conventional analog BF techniques (e.g., \cite{Mo:BF:17}). 
Then the $l$-th CIR tap of the effective mmWave channel is given by 
\begin{align}\label{eq:equi_ch}
	{\bf H}[\ell] &= ({\bf F}_{\rm rx}^{\rm RF})^{\sf H}{\bf A}[\ell] {\bf F}_{\rm tx}^{\rm RF},
\end{align}
for $\ell\in\{0,\ldots,L-1\}$. In this representation, the effects of the antenna array, the transmit analog BF, and the receive analog BF are abstracted by the channel coefficients in $\{{\bf H}[\ell]\}_{\ell}$. Extensive studies and measurement evidences have already shown that the CIR taps of the mmWave channel are sparsely distributed in the delay domain \cite{Mo:18,Akdeniz:14,Rappaport:15,Samimi:16}, because the vulnerability of mmWave signals to reflection and diffraction effects significantly decreases the number of \textit{effective} channel paths between the transmitter and the receiver. Motivated by this fact, we denote the set of non-zero CIR taps as $\mathcal{S}=\{\ell \mid {\bf H}[\ell] \neq {\bf 0}\}$ which is expected to satisfy  $|\mathcal{S}| \ll L$ by the delay-domain sparsity in the mmWave channels.
%Then the delay-domain sparsity in the mmWave channels implies that $|\mathcal{S}| \ll L$. 

\vspace{1mm}
{\bf Numerical example (Delay-domain sparsity in mmWave channels):} 
	We also demonstrate the delay-domain sparsity in mmWave channels by a numerical example using a measurement model in \cite{Samimi:16}. Fig.~\ref{fig:TapDist}(a) plots the cumulative distribution function (CDF) of $L$ and $|\mathcal{S}|$ for various mmWave channels implemented\footnote{In this implementation, the system bandwidth is set to be 1 GHz, the transmitter is assumed to use $4\times 4$ uniform-planar-array (UPA) with $N_{\rm tx}=1$ RF chain, and the receiver is assumed to use $8\times 8$ UPA with $N_{\rm rx}=8$ RF chains. The antenna-element spacing in both the horizontal and the vertical domains of the UPA is set to be $0.5\lambda$. The transmit and receive analog BFs are designed based on Algorithm~1 in \cite{Mo:BF:17}.} from \cite{Samimi:16}, while Fig.~\ref{fig:TapDist}(b) plots a typical power-delay distribution of a 28-GHz non-line-of-sight (NLOS) channel. Both Figs.~\ref{fig:TapDist}(a) and~\ref{fig:TapDist}(b) show that the number of non-zero CIR taps is significantly less than the maximum delay index, i.e., $|\mathcal{S}|\ll L$. These numerical results support our previous discussion on the delay-domain sparsity of the mmWave channels. Fig.~\ref{fig:TapDist}(b) also shows that some non-zero CIR taps have very large discrete-time delays in the range of $450\sim 500$. The reason is that the delay index is a relative value of the propagation delay to the symbol duration; thereby, the larger the system bandwidth, the higher the delay index for the given propagation delay.

%%%%%%%%%%%%%%%%%%%%%%%%%%%%%%%%%%%%%%%%%%%%%%%%%%%%%%%%%%%%%%%%%%%%%%%%%%%%%%%%%%%%%%%%%%%%
%%%%%%%%%%%%%%%%%%%%%%%%%%%%%%%%%%%%%%%%%%%%%%%%%%%%%%%%%%%%%%%%%%%%%%%%%%%%%%%%%%%%%%%%%%%%
\subsection{Signal Model}
At the transmitter, $I_{\rm info}$ information bits intended to be sent to the receiver are encoded into $I_{\rm code}$ coded bits by a channel encoder. Then every group of $M$ coded bits is modulated into an $N_{\rm tx}$-dimensional symbol vector by a symbol mapper.
%The transmitter intends to send $I_{\rm info}$ information bits to the receiver. These information bits are encoded into $I_{\rm code}$ coded bits by a channel encoder, then every group of $M$ coded bits is modulated into an $N_{\rm tx}$-dimensional symbol vector by a symbol mapper. 
%The modulated $N_{\rm d}=\frac{I_{\rm code}}{M}$ symbol vectors are sequentially transmitted over $N_{\rm d}$ time slots, in a manner that the $t$-th element of the symbol vector is transmitted by the $t$-th RF chain at each time slot. 
A symbol vector transmitted at time slot $n$ is denoted by ${\bf x}[n]\in \mathcal{X}$ for $n\in\{1,\ldots,N_{\rm d}\}$, where $N_{\rm d}=\frac{I_{\rm code}}{M}$ and $\mathcal{X}$ is the modulation set for the symbol vector with $|\mathcal{X}|=2^M$. The modulation set is assumed to satisfy the power constraint of $\mathbb{E}[|{x}_t[n]|^2]=1$ for all $t\in\{1,\ldots,N_{\rm tx}\}$, where ${x}_t[n]$ is the $t$-th element of ${\bf x}[n]$. For example, if 4-QAM modulation is used in each RF chain, the symbol vector set is given by 
\begin{align}
	\mathcal{X}=\bigg\{\frac{1}{\sqrt{2}}+j\frac{1}{\sqrt{2}},-\frac{1}{\sqrt{2}}+j\frac{1}{\sqrt{2}},\frac{1}{\sqrt{2}}-j\frac{1}{\sqrt{2}},-\frac{1}{\sqrt{2}}-j\frac{1}{\sqrt{2}}\bigg\}^{N_{\rm tx}}.
\end{align}
Each symbol vector is precoded by the transmit digital BF matrix ${\bf F}_{\rm tx}^{\rm BB}\in\mathbb{C}^{N_{\rm tx}\times N_{\rm tx}}$ with the average power constraint of ${\sf Trace}({\bf F}_{\rm tx}^{\rm BB}\mathbb{E}\left[{\bf x}[n]{\bf x}^{\sf H}[n]\right]({\bf F}_{\rm tx}^{\rm BB})^{\sf H})={N_{\rm tx}}$. Then, using the sparse channel property of the mmWave system, the received signal vector at time slot $n$ is expressed as
\begin{align}\label{eq:received_sig}
	{\bf r}[n]
	&=\sum_{\ell =0}^{L-1}{\bf H}[\ell]{\bf F}_{\rm tx}^{\rm BB}{\bf x}[n-\ell]+ {\bf v}[n] \nonumber \\
	&=\sum_{\ell \in \mathcal{S}}{\bf H}[\ell]{\bf F}_{\rm tx}^{\rm BB}{\bf x}[n-\ell]+ {\bf v}[n], 
\end{align}
where ${\bf v}[n]=\big[v_1[n],\cdots,v_{N_{\rm rx}}[n]\big]^\top\sim\mathcal{CN}({\bf 0}_{N_{\rm rx}},{\sigma^2}{\bf I}_{N_{\rm rx}})$ is a complex Gaussian noise vector. 
%The obtained $N_{\rm d}$ precoded symbol vectors are sequentially transmitted over $N_{\rm d}$ time slots, in a manner that the $t$-th element of the precoded vector is transmitted by the $t$-th RF chain. 

At the ADCs, the real and imaginary parts of each element of ${\bf r}[n]$ are separately quantized by two $B$-bit scalar quantizers. Let $Q: \mathbb{R} \rightarrow \mathcal{Q}=\{q_1,q_2,\ldots,q_{2^B}\}$ be the quantization function of each scalar quantizer, defined as $Q(r)=q_p$ for $r\in\mathbb{R}$ if $b_{p-1} \!<\! r \!\leq\! b_{p}$, where  $q_p$ is the $p$-th quantization output, and $b_p$ is the $p$-th quantization bin boundary such that $b_0\!=\!-\infty \!<\! b_1 \!<\! \ldots \!<\! b_{2^{B-1}} \!<\! b_{2^{B}}\!=\!\infty$. Using this function, the quantized signal vector obtained after the ADCs at time slot $n$ is defined as
\begin{align}\label{eq:quantized_sig}
	{\bf y}[n] = Q\left({\sf Re}\big\{{\bf r}[n]\big\}\right) 
	+ j Q\left({\sf Im}\big\{{\bf r}[n]\big\}\right).
\end{align}
The sequence of the quantized vector obtained during ${N}_{\rm d}+L-1$ time slots is denoted by ${\bf Y}\!=\!\big({\bf y}[1],\cdots,{\bf y}[{N}_{\rm d}\!+\!L\!-\!1]\big)$, which is used as an input of the soft-output detection. Note that in the hybrid BF architecture, the quantized signal in \eqref{eq:quantized_sig} can be combined by a receive digital BF matrix ${\bf F}_{\rm rx}^{\rm BB}\in\mathbb{C}^{N_{\rm rx}\times N_{\rm rx}}$ before the soft-output detection, as illustrated in Fig.~\ref{fig:System}. This additional process, however, cannot improve the performance of the subsequent soft-output detection, because linearly combining the quantized signals only maintains or loses the amount of the information that can be used in the detection. Since the focus of our work is to develop the soft-output detection methods by optimally combining the quantized observations in the detection method, we assume a simple digital BF at the receiver (i.e., ${\bf F}_{\rm rx}^{\rm BB}={\bf I}_{N_{\rm rx}}$).

%%%%%%%%%%%%%%%%%%%%%%%%%%%%%%%%%%%%%%%%%%%%%%%%%%%%%%%%%%%%%%%%%%%%%%%%%%%%%%%%%%%%%%%%%%%%
%%%%%%%%%%%%%%%%%%%%%%%%%%%%%%%%%%%%%%%%%%%%%%%%%%%%%%%%%%%%%%%%%%%%%%%%%%%%%%%%%%%%%%%%%%%%
\subsection{Soft-Output Detection}
The goal of the soft-output detection is to produce the \textit{log-likelihood ratios (LLRs)} of all coded bits based on the quantized observations, so that they can be used as an input of a soft-input channel decoder. Let $c[i]$ be the $i$-th coded bit (i.e., the $i$-th bit output of the channel encoder). In the mmWave MIMO systems with low-precision ADCs, the LLR of the $i$-th coded bit is defined as
\begin{align}\label{eq:LLR_def}
	\mathcal{L}[i]\!=\!\log\frac{\mathbb{P}(c[i]=0|{\bf Y})}{\mathbb{P}(c[i]=1|{\bf Y})},~\text{for}~i\in\{1,\ldots,I_{\rm code}\},
\end{align}
for the given sequence of the quantized received vector, ${\bf Y}$. The above LLR can be rewritten as a function of the a posteriori probability (APP) of the transmitted symbol vector, denoted by $\mathbb{P}({\bf x}[n]|{\bf Y})$. To show this, let $\mathcal{K}_{m}(u)$ be a set of symbol vector indexes that obtain bit $u$ as its $m$-th bit output after a symbol-vector demapping, defined as
\begin{align}\label{eq:X_demap}
	\mathcal{K}_{m}(u) = \big\{ k ~\big|~ {\sf Demap}_m({\bf x}_k)=u, k\in\mathcal{K} \big\},~~\text{for}~~u\in\{0,1\},
\end{align}
where ${\bf x}_k$ is the $k$-th element of $\mathcal{X}$, $\mathcal{K}=\{1,\ldots,|\mathcal{X}|=2^M\}$, and ${\sf Demap}_m(\cdot):\mathcal{X} \rightarrow \{0,1\}$ is the $m$-th bit output of a symbol-vector demapping function. Then the LLR in \eqref{eq:LLR_def} is rewritten as
\begin{align}
	\mathcal{L}[i]
    &= \log\frac{\sum_{k \in \mathcal{K}_{m_i}(0)} \mathbb{P}({\bf x}[n_i]={\bf x}_k|{\bf Y})}{\sum_{k \in \mathcal{K}_{m_i}(1)} \mathbb{P}({\bf x}[n_i]={\bf x}_k|{\bf Y})} \label{eq:LLR_eq1} \\
    &= \log\frac{\sum_{k \in \mathcal{K}_{m_i}(0)} \mathbb{P}({\bf x}[n_i]={\bf x}_k,{\bf Y})}
		{\sum_{k \in \mathcal{K}_{m_i}(1)} \mathbb{P}({\bf x}[n_i]={\bf x}_k,{\bf Y})},  \label{eq:LLR_eq2} 
\end{align}
for $i\in\{1,\ldots,I_{\rm code}\}$, where $n_i=\lceil\frac{i}{M}\rceil$ and $m_i=i-M(n_i-1)$. Note that $n_i$ and $m_i$ are defined in a way that ${\sf Demap}_{m_i}({\bf x}[n_i])=c[i]$ for all $i\in\{1,\ldots,I_{\rm code}\}$. As can be seen in \eqref{eq:LLR_eq1} and \eqref{eq:LLR_eq2}, the LLRs of the coded bits can be determined either by computing the APP of $\mathbb{P}({\bf x}[n]={\bf x}_k|{\bf Y})$, or by computing the marginal  probability of $\mathbb{P}({\bf x}[n]={\bf x}_k,{\bf Y})$, for all $n\in\{1,\ldots,N_{\rm d}\}$ and $k\in\mathcal{K}$.

%%%%%%%%%%%%%%%%%%%%%%%%%%%%%%%%%%%%%%%%%%%%%%%%%%%%%%%%%%%%%%%%%%%%%%%%%%%%%%%%%%%%%%%%%%%%
%%%%%%%%%%%%%%%%%%%%%%%%%%%%%%%%%%%%%%%%%%%%%%%%%%%%%%%%%%%%%%%%%%%%%%%%%%%%%%%%%%%%%%%%%%%%
%%%%%%%%%%%%%%%%%%%%%%%%%%%%%%%%%%%%%%%%%%%%%%%%%%%%%%%%%%%%%%%%%%%%%%%%%%%%%%%%%%%%%%%%%%%%
\section{Construction of Extremely Sparse ISI Channel}
One intriguing aspect of using the low-precision ADCs at the receiver is that it is possible to model the sparse mmWave channel as an extremely sparse ISI channel using the fact that the quantization noise level is sufficiently high. Particularly, under high quantization noise, treating some weak ISI signals as additional noise may not severely degrade the FER performance, while reducing the computational complexity of the soft-output detection. Motivated by this fact, in this section, we first construct an extremely sparse ISI channel model for the mmWave system with low-precision ADCs, which consists only of a few \textit{dominant} CIR taps. We then optimize the selection of the dominant CIR taps to reduce the modeling error in the extremely sparse ISI channel.

%%%%%%%%%%%%%%%%%%%%%%%%%%%%%%%%%%%%%%%%%%%%%%%%%%%%%%%%%%%%%%%%%%%%%%%%%%%%%%%%%%%%%%%%%%%%
%%%%%%%%%%%%%%%%%%%%%%%%%%%%%%%%%%%%%%%%%%%%%%%%%%%%%%%%%%%%%%%%%%%%%%%%%%%%%%%%%%%%%%%%%%%%
\subsection{Extremely Sparse ISI Channel}
Let $\mathcal{D}=\{d_1,d_2,\ldots,d_{|\mathcal{D}|}\}\subset\mathcal{S}$ be a subset of $\mathcal{S}$ that consists of the delays of the dominant CIR taps, and also let $\mathcal{W}=\mathcal{S}/{\mathcal D}=\{w_1,w_2,\ldots,w_{|\mathcal{W}|}\}$ be the non-overlapping subset that consists of the delays of weak CIR taps which are not selected as the dominant CIR taps. Using these two non-overlapping subsets, we rewrite the receive signal in \eqref{eq:received_sig} as
\begin{align}\label{eq:received_dom}
	{\bf r}[n]
	&= \sum_{\ell\in\mathcal{D}}{\bf H}[\ell]{\bf F}_{\rm tx}^{\rm BB}{\bf x}[n-\ell] + \sum_{\ell\in\mathcal{W}}{\bf H}[\ell]{\bf F}_{\rm tx}^{\rm BB}{\bf x}[n-\ell] + {\bf v}[n]  \nonumber \\
	&= {\bf H}_{\mathcal D}{\bf x}_{\mathcal D}[n] + {\bf H}_{\mathcal W}{\bf x}_{\mathcal W}[n] + {\bf v}[n],
\end{align}
where ${\bf H}_{\mathcal D}= \big[{\bf H}[d_{1}]{\bf F}_{\rm tx}^{\rm BB},\cdots,{\bf H}[d_{|\mathcal{D}|}]{\bf F}_{\rm tx}^{\rm BB}\big]$, ${\bf H}_{\mathcal W} = \big[{\bf H}[w_{1}]{\bf F}_{\rm tx}^{\rm BB},\cdots,{\bf H}[w_{|\mathcal{W}|}]{\bf F}_{\rm tx}^{\rm BB}\big]$,
\begin{align}
	{\bf x}_{\mathcal D}[n] = \left[\!\!\begin{array}{c}
	{\bf x}[n\!-\!d_{1}]  \\ \vdots \\ {\bf x}[n\!-\!d_{|\mathcal{D}|}]
	\end{array}\!\!\right],~~\text{and}~~
	{\bf x}_{\mathcal W}[n]= \left[\!\!\begin{array}{c}
	{\bf x}[n\!-\!w_{1}]  \\ \vdots \\ {\bf x}[n\!-\!w_{|\mathcal{W}|}]
	\end{array}\!\!\right].
\end{align} 
In \eqref{eq:received_dom}, the ISI power from the weak CIR taps can be made sufficiently low compared to the quantization noise level by properly determining $\mathcal{D}$ and $\mathcal{W}$. Using this fact, we treat the ISI signals from the weak CIR taps as additional Gaussian noise by modeling ${\bf x}[n-w_{i}]\sim\mathcal{CN}({\bf 0}_{N_{\rm rx}},{\bf I}_{N_{\rm rx}})$ for $w_i\in\mathcal{W}$. Then we can approximately model the effective noise vector $\tilde{\bf v}[n]= \tilde{\bf H}_{\rm W}\tilde{\bf x}_{\mathcal W}[n] + {\bf v}[n]$ in \eqref{eq:received_dom} as a complex Gaussian random vector with zero-mean and the covariance matrix of 
\begin{align}\label{eq:Cov_v}
	\mathbb{E}\left[ \tilde{\bf v}[n] \tilde{\bf v}^{\sf H}[n]\right] 
	&= {\bf H}_{\mathcal W} \mathbb{E}\left[ {\bf x}_{\mathcal W}[n]{\bf x}_{\mathcal W}^{\sf H}[n]\right] {\bf H}_{\rm W}^{\sf H} + {\sigma^2}{\bf I}_{N_{\rm rx}} \nonumber \\
	&= {\bf H}_{\mathcal W}^{(n)} \big({\bf H}_{\rm W}^{(n)}\big)^{\sf H} + {\sigma^2}{\bf I}_{N_{\rm rx}},
	%	&\overset{(a)}{\simeq} {\sf diag}\left(\tilde\sigma_{1}^2[n],\ldots, \tilde\sigma_{N_{\rm rx}}^2[n]\right),
\end{align}
where ${\bf H}_{\mathcal W}^{(n)}$ is a sub-matrix of ${\bf H}_{\mathcal W}$ that only contains the weak CIR taps associating with non-zero transmitted vectors at time slot $n$. Since ${\bf H}_{\mathcal W}^{(n)} \big({\bf H}_{\rm W}^{(n)}\big)^{\sf H}$ is a diagonal dominant matrix for the spatially uncorrelated MIMO channel environment, we further approximate the covariance of the effective noise as 
\begin{align}\label{eq:Cov_approx}
	\mathbb{E}\left[ \tilde{\bf v}[n] \tilde{\bf v}^{\sf H}[n]\right] 
	\approx {\sf diag}\big(\tilde\sigma_{1}^2[n],\ldots, \tilde\sigma_{N_{\rm rx}}^2[n]\big),
\end{align}	
where $\tilde{\sigma}_{r}^2[n] = \|{\bf h}_{{\mathcal W},r}^{(n)}\|^2 + \sigma^2$, and $({\bf h}_{{\mathcal W},r}^{(n)})^\top$ is the $r$-th row of ${\bf H}_{\mathcal W}^{(n)}$.
Our effective noise model can be made accurate by selecting the weak CIR taps whose sum power is much lower than the noise level, i.e.,  $\|{\bf h}_{{\mathcal W},r}^{(n)}\|^2 \ll \sigma^2$ for $r\in\{1,\ldots,N_{\rm rx}\}$. The selection method will be explained in the following subsection.
By applying the effective noise model, we rewrite the quantized received vector in \eqref{eq:quantized_sig} as 
\begin{align}\label{eq:quantized_dom}
	{\bf y}[n] = Q\left({\sf Re}\Big\{{\bf H}_{\mathcal D}{\bf x}_{\mathcal D}[n] + \tilde{\bf v}[n]\Big\}\right) 
	+ j Q\left({\sf Im}\Big\{{\bf H}_{\mathcal D}{\bf x}_{\mathcal D}[n] + \tilde{\bf v}[n]\Big\}\right).
\end{align}
As seen in \eqref{eq:quantized_dom}, the quantized received signal, ${\bf y}[n]$, can be effectively modeled as the quantized output of the extremely sparse ISI channel with independent colored noise. It is also noticeable that in the above model, the quantized received signal for $n\geq  N_{\rm d}+L_{\mathcal D}$ is ignored,  where $L_{\mathcal D}=\max_{\ell \in\mathcal{D}}\ell + 1$ is the maximum delay index of the dominant CIR taps. Therefore, only the partial sequence $\tilde{\bf Y}=\big({\bf y}[1],\cdots,{\bf y}[{N}_{\rm d}+L_{\mathcal D}-1]\big)$ is used in the soft-output detection, instead of the full sequence ${\bf Y}=\big({\bf y}[1],\cdots,{\bf y}[{N}_{\rm d}+L-1]\big)$.

We also characterize the conditional probability mass function (PMF) of the constructed sparse ISI channel, which will be harnessed as the sufficient statistic for the soft-output detection. From \eqref{eq:quantized_dom}, the conditional PMF of observing ${\bf y}[n]$ for given ${\bf x}_{\mathcal D}[n]$ is approximately computed as
\begin{align}\label{eq:condPMF_dom}
	&\mathbb{P}({\bf y}[n]|{\bf x}_{\mathcal D}[n])  \nonumber \\
	&= \prod_{r=1}^{N_{\rm rx}}
	\mathbb{P}\Big( l({y}_r^{\rm Re}[n]) < {\sf Re}\big\{{\bf h}_{{\mathcal D},r}{\bf x}_{\mathcal D}[n] + \tilde{v}_r[n]\big\} \leq u({y}_r^{\rm Re}[n]) \Big) \nonumber \\
	&~~~~~~~~\times \mathbb{P}\Big( l({y}_r^{\rm Im}[n]) < {\sf Im}\big\{{\bf h}_{{\mathcal D},r}^\top{\bf x}_{\mathcal D}[n] + \tilde{v}_r[n]\big\} \leq u({y}_r^{\rm Im}[n]) \Big) \nonumber \\
	&= \prod_{r=1}^{N_{\rm rx}}	
	\Bigg[\Phi \Bigg(	\frac{u({y}_r^{\rm Re}[n])- {\sf Re}\big\{{\bf h}_{{\mathcal D},r}^\top{\bf x}_{\mathcal D}[n]\big\}}{\sqrt{(\sigma^2+\|{\bf h}_{{\mathcal W},r}^{(n)}\|^2)/2}}\Bigg)
	-\Phi\Bigg(\frac{l({y}_r^{\rm Re}[n])-{\sf Re}\big\{{\bf h}_{{\mathcal D},r}^\top{\bf x}_{\mathcal D}[n]\big\}}{\sqrt{(\sigma^2+\|{\bf h}_{{\mathcal W},r}^{(n)}\|^2)/2}}\Bigg)\Bigg] \nonumber \\
	&~~~~~~~~\times \Bigg[\Phi \Bigg(	\frac{u({y}_r^{\rm Im}[n])-{\sf Im}\big\{{\bf h}_{{\mathcal D},r}^\top{\bf x}_{\mathcal D}[n]\big\}}{\sqrt{(\sigma^2+\|{\bf h}_{{\mathcal W},r}^{(n)}\|^2)/2}}\Bigg)
	-\Phi\Bigg(\frac{l({y}_r^{\rm Im}[n])-{\sf Im}\big\{{\bf h}_{{\mathcal D},r}^\top{\bf x}_{\mathcal D}[n]\big\}}{\sqrt{(\sigma^2+\|{\bf h}_{{\mathcal W},r}^{(n)}\|^2)/2}}\Bigg)\Bigg],
\end{align}
where ${\bf h}_{{\mathcal D},r}^\top$ is the $r$-th row of ${\bf H}_{\mathcal D}$, ${y}_r^{\rm Re}[n]={\sf Re}\{y_r[n]\}$, ${y}_r^{\rm Im}[n]={\sf Im}\{y_r[n]\}$, and $l(q_p)=b_{p-1}$ and $u(q_p)=b_p$ are the lower and upper quantization boundaries associating with $q_p\in\mathcal{Q}$, respectively. Since the conditional PMF in \eqref{eq:condPMF_dom} is computed based on the approximate model in \eqref{eq:quantized_dom}, it differs from the \textit{true} conditional PMF that does not treat the ISI signals from weak CIR taps as noise, given by
\begin{align}\label{eq:condPMF_true}
	\mathbb{P}({\bf y}[n]|{\bf x}_{\mathcal D}[n],{\bf x}_{\mathcal W}[n])
	&=\prod_{r=1}^{N_{\rm rx}}
	\Bigg[\Phi \Bigg(	\frac{u({y}_r^{\rm Re}[n])-{\sf Re}\big\{{\bf h}_{{\mathcal D},r}^\top{\bf x}_{\mathcal D}[n]+{\bf h}_{{\mathcal W},r}^\top{\bf x}_{\mathcal W}[n]\big\}}{\sqrt{\sigma^2/2}}\Bigg)   \nonumber \\
	&~~~~~~~~~~~-\Phi\Bigg(\frac{l({y}_r^{\rm Re}[n])-{\sf Re}\big\{{\bf h}_{{\mathcal D},r}^\top{\bf x}_{\mathcal D}[n]+{\bf h}_{{\mathcal W},r}^\top{\bf x}_{\mathcal W}[n]\big\}}{\sqrt{\sigma^2/2}}\Bigg)\Bigg] \nonumber \\
	&~~~~~\times\Bigg[\Phi \Bigg(	\frac{u({y}_r^{\rm Im}[n])-{\sf Im}\big\{{\bf h}_{{\mathcal D},r}^\top{\bf x}_{\mathcal D}[n]+{\bf h}_{{\mathcal W},r}^\top{\bf x}_{\mathcal W}[n]\big\}}{\sqrt{\sigma^2/2}}\Bigg)   \nonumber \\
	&~~~~~~~~~~~-\Phi\Bigg(\frac{l({y}_r^{\rm Im}[n])-{\sf Im}\big\{{\bf h}_{{\mathcal D},r}^\top{\bf x}_{\mathcal D}[n]+{\bf h}_{{\mathcal W},r}^\top{\bf x}_{\mathcal W}[n]\big\}}{\sqrt{\sigma^2/2}}\Bigg)\Bigg],
\end{align}
where ${\bf h}_{{\mathcal W},r}^\top$ is the $r$-th row of ${\bf H}_{\mathcal W}$. The comparison between \eqref{eq:condPMF_dom} and \eqref{eq:condPMF_true} reveals that the use of the extremely sparse ISI channel in \eqref{eq:quantized_dom} causes a mismatch in the conditional PMF, which can be interpreted as a modeling error.

%%%%%%%%%%%%%%%%%%%%%%%%%%%%%%%%%%%%%%%%%%%%%%%%%%%%%%%%%%%%%%%%%%%%%%%%%%%%%%%%%%%%%%%%%%%%
%%%%%%%%%%%%%%%%%%%%%%%%%%%%%%%%%%%%%%%%%%%%%%%%%%%%%%%%%%%%%%%%%%%%%%%%%%%%%%%%%%%%%%%%%%%%
\subsection{Dominant-Tap-Selection Algorithm}
To reduce the modeling error in the extremely sparse ISI channel, we develop a dominant-tap-selection algorithm that minimizes the mismatch between the approximate conditional PMF in \eqref{eq:condPMF_dom} and the true conditional PMF in \eqref{eq:condPMF_true}. One simple solution to achieve this goal is to exhaustively search for the best dominant CIR taps that minimize the Kullback-Leibler (KL) divergence between two conditional PMFs, defined as
\begin{align}\label{eq:KL_diverse}
D_{\rm KL}(\mathcal{D}) 
= \sum_{{\bf y}\in\{\mathcal{Q}^{N_{\rm rx}}+j\mathcal{Q}^{N_{\rm rx}}\}} \mathbb{P}({\bf y}|{\bf x}_{\mathcal D}) \ln\left(\frac{\mathbb{P}({\bf y}|{\bf x}_{\mathcal D})}{\mathbb{P}({\bf y}|{\bf x}_{\mathcal D},{\bf x}_{\mathcal W})}\right),
\end{align}
for ${\bf x}_{\mathcal D}\in\mathcal{X}^{|\mathcal{D}|}$ and ${\bf x}_{\mathcal W}\in\mathcal{X}^{|\mathcal{W}|}$. Unfortunately, computing the KL divergence in \eqref{eq:KL_diverse} for all possible  ${\bf x}_{\mathcal D}\in\mathcal{X}^{|\mathcal{D}|}$ and ${\bf x}_{\mathcal W}\in\mathcal{X}^{|\mathcal{W}|}$ requires a prohibitive computational complexity that increases with the size of the input sets (i.e., $|\mathcal{X}|^{|\mathcal{D}|}$ and $|\mathcal{X}|^{|\mathcal{W}|}$) and also requires the complicated computations of the conditional PMFs. Therefore, to avoid this difficulty, we focus on developing a computationally-efficient algorithm that operates with a closed-form criterion.
%evelop a greedy algorithm based on a closed-form criterion that significantly reduces the computational complexity required for the dominant-tap-selection process. 

%To obtain the closed-form criterion, our strategy is to reduce the difference between the arguments of the true and approximate conditional PMFs, instead of directly minimizing the difference between these two PMFs. Based on this strategy, we design a normalized mean-squared-error (NMSE) criterion measured between the arguments of two PMFs. 

We start by designing a closed-form criterion for the dominant-tap selection. The key idea is to reduce the difference between the \textit{arguments} of the true and approximate conditional PMFs, instead of directly minimizing the difference between the PMFs. Based on this idea, we adopt a normalized mean-squared-error (NMSE) criterion measured between the arguments of the true and approximate conditional PMFs. Let $\phi_{p,r}^{\rm Re}({\bf x}_{\mathcal D},{\bf x}_{\mathcal W}; \mathcal{D})$ and $\phi_{p,r}^{\rm Im}({\bf x}_{\mathcal D},{\bf x}_{\mathcal W}; \mathcal{D})$ be the arguments of the true PMF in \eqref{eq:condPMF_true}:
\begin{align*}
	\phi_{p,r}^{\rm Re}({\bf x}_{\mathcal D},{\bf x}_{\mathcal W}; \mathcal{D})
	&= \frac{b_p -{\sf Re}\big\{{\bf h}_{{\mathcal D},r}^\top{\bf x}_{\mathcal D} + {\bf h}_{{\mathcal W},r}^\top{\bf x}_{\mathcal W}\big\}}{\sqrt{\sigma^2/2}}, \\
	\phi_{p,r}^{\rm Im}({\bf x}_{\mathcal D},{\bf x}_{\mathcal W}; \mathcal{D})
	&= \frac{b_p -{\sf Im}\big\{{\bf h}_{{\mathcal D},r}^\top{\bf x}_{\mathcal D} + {\bf h}_{{\mathcal W},r}^\top{\bf x}_{\mathcal W}\big\}}{\sqrt{\sigma^2/2}}, 
\end{align*}
for $r\in\{1,\ldots,N_{\rm rx}\}$ and $p\in\{0,\ldots,2^B\}$. Similarly, let $\hat{\phi}_{p,r}^{\rm Re}({\bf x}_{\mathcal D}; \mathcal{D})$ and $\hat{\phi}_{p,r}^{\rm Im}({\bf x}_{\mathcal D}; \mathcal{D})$ be the arguments of the approximate PMF in \eqref{eq:condPMF_dom}:
\begin{align*}
	\hat{\phi}_{p,r}^{\rm Re}({\bf x}_{\mathcal D}; \mathcal{D})
	&= \frac{b_p -{\sf Re}\big\{{\bf h}_{{\mathcal D},r}^\top{\bf x}_{\mathcal D}\big\}}{\sqrt{(\sigma^2 + \|{\bf h}_{{\mathcal W},r}\|^2)/2}}, \nonumber \\
	\hat{\phi}_{p,r}^{\rm Im}({\bf x}_{\mathcal D}; \mathcal{D})
	&= \frac{b_p -{\sf Im}\big\{{\bf h}_{{\mathcal D},r}^\top{\bf x}_{\mathcal D}\big\}}{\sqrt{(\sigma^2 + \|{\bf h}_{{\mathcal W},r}\|^2)/2}},
\end{align*}
for $r\in\{1,\ldots,N_{\rm rx}\}$ and $p\in\{0,\ldots,2^B\}$. Then the NMSE between the above arguments is defined as 
\begin{align}\label{eq:NMSE_def}
	{\sf NMSE}(\mathcal{D})
	= \sum_{r=1}^{N_{\rm rx}}\sum_{p=1}^{2^B-1}\frac{1}{2}
	&\Bigg\{\frac{\mathbb{E}\big[|\phi_{p,r}^{\rm Re}({\bf x}_{\mathcal D},{\bf x}_{\mathcal W}; \mathcal{D}) - \hat\phi_{p,r}^{\rm Re}({\bf x}_{\mathcal D}; \mathcal{D}) |^2 \big]}
	{\mathbb{E}\big[|\hat\phi_{p,r}^{\rm Re}({\bf x}_{\mathcal D}; \mathcal{D})|^2\big]} \nonumber \\
	&~~~~+ \frac{\mathbb{E}\big[|\phi_{p,r}^{\rm Im}({\bf x}_{\mathcal D},{\bf x}_{\mathcal W}; \mathcal{D}) - \hat\phi_{p,r}^{\rm Im}({\bf x}_{\mathcal D}; \mathcal{D}) |^2 \big]}
	{\mathbb{E}\big[|\hat\phi_{p,r}^{\rm Im}({\bf x}_{\mathcal D}; \mathcal{D})|^2\big]}\Bigg\},
\end{align}
where the expectation is with respect to ${\bf x}_{\mathcal D}$ and ${\bf x}_{\mathcal W}$. Note that we use the NMSE instead of the MSE to incorporate different MSEs equally contribute to the sum of them.	To obtain a closed-form expression for the NMSE criterion, we further model ${\bf x}_{\mathcal D}$ and ${\bf x}_{\mathcal W}$ as complex Gaussian signals distributed as $\mathcal{CN}({\bf 0}_{|\mathcal{D}|N_{\rm tx}},{\bf I}_{|\mathcal{D}|N_{\rm tx}})$ and $\mathcal{CN}({\bf 0}_{|\mathcal{W}|N_{\rm tx}},{\bf I}_{|\mathcal{W}|N_{\rm tx}})$, respectively. Under the Gaussian modeling, we obtain the following distributions: 
\begin{align}
	{\sf Re}\big\{{\bf h}_{{\mathcal D},r}^\top{\bf x}_{\mathcal D}\big\},{\sf Im}\big\{{\bf h}_{{\mathcal D},r}^\top{\bf x}_{\mathcal D}\big\}  
	&\sim \mathcal{N}\big(0,{P_{{\rm D},r}(\mathcal{D})}/{2}\big), \nonumber \\
	{\sf Re}\big\{{\bf h}_{{\mathcal W},r}^\top{\bf x}_{\mathcal W}\big\},{\sf Im}\big\{{\bf h}_{{\mathcal W},r}^\top{\bf x}_{\mathcal W}\big\}  
	&\sim \mathcal{N}\big(0,{P_{{\rm W},r}(\mathcal{D})}/{2}\big), \nonumber 
\end{align}
where $P_{{\rm D},r}(\mathcal{D})=\|{\bf h}_{\mathcal{D},r}\|^2$, and $P_{{\rm W},r}(\mathcal{D})=\|{\bf h}_{\mathcal{W},r}\|^2$. Using this fact, we compute the expectation terms in \eqref{eq:NMSE_def} as
\begin{align}\label{eq:MSE_nom}
	&\mathbb{E}_{{\bf x}_{\mathcal D},{\bf x}_{\mathcal W}}\big[|\phi_{p,r}^{\rm Re}({\bf x}_{\mathcal D},{\bf x}_{\mathcal W}; \mathcal{D}) 
	- \hat\phi_{p,r}^{\rm Re}({\bf x}_{\mathcal D}; \mathcal{D}) |^2 \big] \nonumber \\
	&= \frac{\mathbb{E}_{{\bf x}_{\mathcal D}}\big[\big(b_p-{\sf Re}\big\{{\bf h}_{{\mathcal D},r}^\top{\bf x}_{\mathcal D}\big\}\big)^2\big]}{\sigma^2/2}
	\left(1 - \sqrt{\frac{\sigma^2 }{\sigma^2 + P_{{\rm W},r}(\mathcal{D})}} \right)^2
	+\frac{\mathbb{E}_{{\bf x}_{\mathcal W}}\big[{\sf Re}\big\{{\bf h}_{{\mathcal W},r}^\top{\bf x}_{\mathcal W}\big\}^2\big]}{\sigma^2/2} \nonumber \\
	&= \frac{2b_p^2+P_{{\rm D},r}(\mathcal{D})}{\sigma^2}
	\left(1 - \sqrt{\frac{\sigma^2 }{\sigma^2 + P_{{\rm W},r}(\mathcal{D})}} \right)^2
	+\frac{P_{{\rm W},r}(\mathcal{D})}{\sigma^2} \nonumber \\
	&=\mathbb{E}_{{\bf x}_{\mathcal D},{\bf x}_{\mathcal W}}\big[|\phi_{p,r}^{\rm Im}({\bf x}_{\mathcal D},{\bf x}_{\mathcal W}; \mathcal{D}) 
	- \hat\phi_{p,r}^{\rm Im}({\bf x}_{\mathcal D}; \mathcal{D}) |^2 \big], 	
\end{align}
and
\begin{align}\label{eq:MSE_denom}
	\mathbb{E}_{{\bf x}_{\mathcal D},{\bf x}_{\mathcal W}}\big[|\hat\phi_{p,r}^{\rm Re}({\bf x}_{\mathcal D}; \mathcal{D})|^2\big]
	=\mathbb{E}_{{\bf x}_{\mathcal D},{\bf x}_{\mathcal W}}\big[|\hat\phi_{p,r}^{\rm Im}({\bf x}_{\mathcal D}; \mathcal{D})|^2\big]
	= \frac{2b_p^2 + P_{{\rm D},r}(\mathcal{D})}{\sigma^2 + P_{{\rm W},r}(\mathcal{D})}.
\end{align}
%\begin{align}\label{eq:MSE_denom}
%	\mathbb{E}_{{\bf x}_{\mathcal D},{\bf x}_{\mathcal W}}\big[|\phi_{p,r}^{\rm Re}({\bf x}_{\mathcal D},{\bf x}_{\mathcal W}; \mathcal{D})|^2\big]
%	=\mathbb{E}_{{\bf x}_{\mathcal D},{\bf x}_{\mathcal W}}\big[|\phi_{p,r}^{\rm Im}({\bf x}_{\mathcal D},{\bf x}_{\mathcal W}; \mathcal{D})|^2\big]
%	= \frac{2b_p^2 + P_{{\rm H},r}}{\sigma^2},
%\end{align}
%where $P_{{\rm H},r} = P_{{\rm D},r}(\mathcal{D}) + P_{{\rm W},r}(\mathcal{D}) = \sum_{\ell=0}^{L-1}\|{\bf h}_{r}[\ell]\|^2$.
By applying \eqref{eq:MSE_nom} and \eqref{eq:MSE_denom} into \eqref{eq:NMSE_def}, we finally obtain the closed-form expression of the NMSE criterion:
\begin{align}\label{eq:NMSE_compute}
	{\sf NMSE}(\mathcal{D})
	&= \sum_{r=1}^{N_{\rm rx}} \sum_{p=1}^{2^B-1} \frac{\sigma^2 + P_{{\rm W},r}(\mathcal{D}) }{\sigma^2 } \Bigg\{
	\Bigg(1-\sqrt{\frac{\sigma^2 }{\sigma^2 + P_{{\rm W},r}(\mathcal{D}) }}\Bigg)^{\!\!2} 
	+  \frac{P_{{\rm W},r}(\mathcal{D})}{2b_p^2 + P_{{\rm D},r}(\mathcal{D})}\Bigg\}.	 
\end{align}
%\begin{align}\label{eq:NMSE_compute}
%	{\sf NMSE}(\mathcal{D})
%	&= \sum_{r=1}^{N_{\rm rx}} \sum_{p=1}^{2^B-1} \Bigg\{
%	\frac{2b_p^2+P_{{\rm D},r}(\mathcal{D})}{2b_p^2 +  P_{{\rm H},r}}
%	\Bigg(1 - \sqrt{\frac{\sigma^2 }{\sigma^2 + P_{{\rm W},r}(\mathcal{D})}} \Bigg)^2 
%	+ \frac{P_{{\rm W},r}(\mathcal{D})}{2b_p^2 +  P_{{\rm H},r}}\Bigg\}.	 
%\end{align}
Although this is not an optimal criterion, it \textit{effectively} reduces the difference between the approximate and true conditional PMFs, because the difference between them strictly decreases as the difference between their arguments decreases. Meanwhile, the use of the NMSE criterion in \eqref{eq:NMSE_compute} significantly reduces the computational complexity of the dominant-tap-selection process, as it requires neither the marginalization over all possible transmitted signals nor the computation of the conditional PMFs. The NMSE criterion also takes into account the effect of the quantization function (i.e., quantization bin boundaries $\{b_p\}_{p=1}^{2^B-1}$). For example, when $\sigma^2 \gg P_{{\rm W},r}(\mathcal{D})$, the quantization boundaries with $b_p^2$ contribute less to the NMSE,  
%decreases with the relative power of the quantization boundaries compared to the input power, 
because these boundaries produce high quantization noises that make the effect of the weak CIR taps to be insignificant.

By harnessing the NMSE criterion in \eqref{eq:NMSE_compute}, we propose a dominant-tap-selection algorithm that finds the best dominant CIR taps with the minimum NMSE using a greedy approach. The proposed algorithm is summarized in Algorithm~\ref{alg:DomTap}. 
\begin{algorithm}[H]
	\caption{The proposed dominant-tap-selection algorithm}\label{alg:DomTap}
	\small{\begin{algorithmic}[1]
			\STATE Initialize $\mathcal{D}^*=\emptyset$ and $\mathcal{W}^*=\mathcal{S}$.
			\WHILE {${\sf NMSE}(\mathcal{D}^*) > \epsilon_{\rm th}$ and $|\mathcal{D}^*| < D_{\rm max}$ and $\mathcal{W}^*\neq \emptyset$}
			\STATE Find $l^* = \argmin_{l\in\mathcal{W}^*}{\sf NMSE}(\mathcal{D}^*\cup\{l\})$ from \eqref{eq:NMSE_compute}.
			\STATE Update $\mathcal{D}^* \leftarrow \mathcal{D}^*\cup\{l^*\}$ and $\mathcal{W}^* \leftarrow \mathcal{W}^*\setminus\{l^*\}$.
			\ENDWHILE
	\end{algorithmic}}
\end{algorithm}

\vspace{-3mm}
\noindent In Algorithm~\ref{alg:DomTap}, we adopt two design parameters: 1) the maximum number of dominant CIR taps, $D_{\rm max}$, and 2) an NMSE threshold, $\epsilon_{\rm th}$. Using these two parameters, the proposed algorithm stops if the NMSE in \eqref{eq:NMSE_compute} falls below the given threshold $\epsilon_{\rm th}$, or if the number of the dominant CIR taps reaches to the maximum number $D_{\rm max}$, or if every CIR tap is selected as the dominant taps (i.e., $\mathcal{W}=\emptyset$). As we will show, these two parameters effectively adjust the performance-complexity tradeoff in the soft-output detection.

%%%%%%%%%%%%%%%%%%%%%%%%%%%%%%%%%%%%%%%%%%%%%%%%%%%%%%%%%%%%%%%%%%%%%%%%%%%%%%%%%%%%%%%%%%%%
%%%%%%%%%%%%%%%%%%%%%%%%%%%%%%%%%%%%%%%%%%%%%%%%%%%%%%%%%%%%%%%%%%%%%%%%%%%%%%%%%%%%%%%%%%%%
%%%%%%%%%%%%%%%%%%%%%%%%%%%%%%%%%%%%%%%%%%%%%%%%%%%%%%%%%%%%%%%%%%%%%%%%%%%%%%%%%%%%%%%%%%%%
\section{Soft-Output Detection Methods for Extremely Sparse ISI Channel}
In this section, based on the extremely sparse ISI channel in Section III, we present two computationally-efficient yet near-optimal algorithms for the soft-output detection in mmWave MIMO systems with low-precision ADCs.

%%%%%%%%%%%%%%%%%%%%%%%%%%%%%%%%%%%%%%%%%%%%%%%%%%%%%%%%%%%%%%%%%%%%%%%%%%%%%%%%%%%%%%%%%%%%
%%%%%%%%%%%%%%%%%%%%%%%%%%%%%%%%%%%%%%%%%%%%%%%%%%%%%%%%%%%%%%%%%%%%%%%%%%%%%%%%%%%%%%%%%%%%
\subsection{Quantized BCJR (Q-BCJR)}
We first develop a soft-output detection method called \textit{quantized BCJR (Q-BCJR)} by modifying the classical BCJR algorithm in \cite{BCJR:74,Li:95} to operate with the quantized outputs over the extremely sparse ISI channel. The basic idea of Q-BCJR is to use a forward-and-backward algorithm based on a reduced trellis-diagram to compute the LLRs of coded bits by optimally combining the quantized received signals obtained from multiple receive antennas. Thanks to the extreme sparsity of the channel, the trellis-diagram of Q-BJCR has a reduced number of states that depend only on the maximum delay of the dominant CIR taps. Particularly, a state vector at time slot $n$ is defined as
\begin{align} \label{eq:def_state}
	{\bf s}[n]=\big[{\bf x}^{\top}[n], {\bf x}^{\top}[n-1], \ldots, {\bf x}^{\top}[n-L_{\mathcal D}+2]\big]^{\top} \in\mathcal{X}^{L_{\mathcal D}-1},
\end{align}
for $n\in\{0,\ldots,N_{\rm d}+L_{\mathcal D}-1\}$.
%where $L_{\mathcal D}=\max_{\ell \in\mathcal{D}}\ell + 1$, and ${\bf x}[n]={\bf 0}_{N_{\rm tx}}$ for $n\!\notin\!\{1,\ldots,{N}_{\rm d}\}$. 
Then the set of valid state vectors at time slot $n$ is defined as
\begin{align} \label{eq:def_state_set}
	\mathcal{S}_n &= \Big\{\big[{\bf x}^{\top}[n], {\bf x}^{\top}[n-1], \ldots, {\bf x}^{\top}[n-L_{\mathcal D}+2]\big]^{\top}\Big| \nonumber \\
	&~~~~~~~~~~{\bf x}[n]\!\in\!\mathcal{X}~\text{for}~n\!\in\!\{1,\ldots,N_{\rm d}\},~ {\bf x}[n]={\bf 0}_{N_{\rm tx}}~\text{for}~n\!\notin\!\{1,\ldots,N_{\rm d}\} \Big\}
\end{align}
By the definition of the state vector, a set of state-vector pair $({\bf s}',{\bf s})$ associating with the event $\{{\bf s}[n-1]={\bf s}',{\bf s}[n]={\bf s},{\bf x}[n]={\bf x}_k\}$ is defined as 
\begin{align} \label{eq:def_state_trans}
	\mathcal{V}_{n,k}
	&=\Big\{({\bf s}',{\bf s})\Big| {\bf s}[n-1]={\bf s}',{\bf s}[n]={\bf s},{\bf x}[n]={\bf x}_k, {\bf s}'\in \mathcal{S}_{n-1},{\bf s}\in\mathcal{S}_n \Big\} \nonumber \\
	&=\Big\{({\bf s}',{\bf s})\Big| {\bf s}=\big[{\bf x}_k^\top, ({\bf s}')_{1:(L_{\mathcal D}-2)N_{\rm tx}}^\top \big]^\top, {\bf s}'\in \mathcal{S}_{n-1},{\bf s}\in\mathcal{S}_n\Big\},
\end{align}
for $k\in\mathcal{K}$ when $1\leq n\leq N_{\rm d}$ and $k=0$ when $N_{\rm d}+1\leq n\leq N_{\rm d}+L_{\mathcal D}-1$, where $({\bf a})_{i:j}=[a_i,a_{i+1},\cdots,a_j]^\top$ is a subvector of ${\bf a}=[a_1,a_2,\cdots,a_N]^\top$ for $i\leq j\leq N$. Note that we define ${\bf x}_0={\bf 0}_{N_{\rm tx}}$ for notational consistency.

Using the above notations, in Q-BCJR, the marginal probability $\mathbb{P}({\bf x}[n]={\bf x}_k,\tilde{\bf Y})$ is factorized into multiple factors as given in the following proposition: 
\begin{prop}\label{Prop1}
	The marginal probability $\mathbb{P}({\bf x}[n]={\bf x}_k,\tilde{\bf Y})$ for $n\in\{1,\ldots,N_{\rm d}\}$ and $k\in\{1,\ldots,2^M\}$ is expressed as 
	\begin{align}\label{eq:marginalProb}
		\mathbb{P}({\bf x}[n]={\bf x}_k,\tilde{\bf Y})  =\sum_{({\bf s}',{\bf s})\in\mathcal{V}_n(k)} \alpha_{n-1}({\bf s}')\gamma_{n}({\bf s}',{\bf s})\beta_{n}({\bf s}),
	\end{align}
	where $\alpha_0({\bf s}) = \mathbb{I}\big\{{\bf s}={\bf 0}_{(L_{\mathcal D}-1)N_{\rm tx}}\big\}$,
	$\beta_{N_{\rm d}+L_{\mathcal D}-1}({\bf s}) =\mathbb{I}\{{\bf s}={\bf 0}_{(L_{\mathcal D}-1)N_{\rm tx}}\}$,
	\begin{align}
	\alpha_{n}({\bf s}) &= \sum_{{\bf s}'\in\mathcal{S}_{n-1}} \alpha_{n-1}({\bf s}')\gamma_n({\bf s}',{\bf s}),
	~~~~~~~~~~~~{\rm for}~n\in\{1,\ldots,N_{\rm d}+L_{\mathcal D}-1\}, \label{eq:alpha} \\  
	\beta_{n-1}({\bf s}') &= \sum_{{\bf s}\in\mathcal{S}_{n}} \beta_n({\bf s})\gamma_n({\bf s}',{\bf s}),
	~~~~~~~~~~~~~~~~~~{\rm for}~n\in\{2,\ldots,N_{\rm d}+L_{\mathcal D}-1\}, \label{eq:beta} \\  
	\gamma_n({\bf s}',{\bf s}) &=\!\!
	\begin{cases}
	\frac{1}{2^M}\mathbb{P}\big({\bf y}[n]\big|{\bf s}[n\!-\!1]\!=\!{\bf s}',{\bf s}[n]\!=\!{\bf s}\big), &\!\!\!\!{\rm for}~({\bf s}',{\bf s})\in\mathcal{V}_n(k), n\in\{1,\ldots,N_{\rm d}\}, \\
	\mathbb{P}\big({\bf y}[n]\big|{\bf s}[n\!-\!1]\!=\!{\bf s}',{\bf s}[n]\!=\!{\bf s}\big), &\!\!\!\!{\rm for}~({\bf s}',{\bf s})\in\mathcal{V}_n(k), n\in\{N_{\rm d}\!+\!1,\ldots,N_{\rm d}\!+\!L_{\mathcal D}\!-\!1\}, \\
	0, &\!\!\!\!{\rm otherwise}.
	\end{cases} \label{eq:gamma} 
	\end{align}
\end{prop}
\begin{IEEEproof}
	See Appendix A.
\end{IEEEproof}

Proposition~1 shows that the marginal probability is determined by three factors, $\alpha_{n-1}({\bf s}')$, $\beta_{n}({\bf s})$, and  $\gamma_{n}({\bf s}',{\bf s})$. As can be seen in \eqref{eq:alpha} and \eqref{eq:beta}, the first two factors $\alpha_{n-1}({\bf s}')$ and $\beta_{n}({\bf s})$ are efficiently computed in a recursive manner. In addition, the remaining factor $\gamma_{n}({\bf s}',{\bf s})$ is directly computed by the approximate model in \eqref{eq:quantized_dom} with \eqref{eq:condPMF_dom} based on the following relation:
\begin{align}\label{eq:condPMF_QBCJR}
	&\mathbb{P}\big({\bf y}[n]\big|{\bf s}[n-1]={\bf s}',{\bf s}[n]={\bf s}\big)   \nonumber \\
	&= \mathbb{P}\Big({\bf y}[n]\Big|\big[{\bf x}^\top[n], \cdots,{\bf x}^{\top}[n-L_{\mathcal D}+1]\big]^\top\!\!
	=\underbrace{\big[({\bf s})_{1:N_{\rm tx}}^\top,({\bf s}')^\top\big]^\top}_{=\bar{\bf x}_n\{{\bf s},{\bf s}'\}}\Big) \nonumber \\
	&= \mathbb{P}\left({\bf y}[n]\Big| {\bf x}_{\mathcal D}[n]
	\!\!=\!\!\left[
	\big(\bar{\bf x}_n\{{\bf s},{\bf s}'\}\big)_{d_{n,1}N_{\rm tx}+1:(d_{n,1}+1)N_{\rm tx}}^\top,  \cdots,
	\big(\bar{\bf x}_n\{{\bf s},{\bf s}'\}\big)_{d_{n,|\mathcal{D}_n|}N_{\rm tx}+1:(d_{n,|\mathcal{D}_n|}+1)N_{\rm tx}}^\top\right] \right).
\end{align}
The computation of $\mathbb{P}\big({\bf y}[n]\big|{\bf s}[n-1]={\bf s}',{\bf s}[n]={\bf s}\big)$ brings the key differences of Q-BCJR to the classical BCJR algorithm \cite{BCJR:74,Li:95}. First, Q-BCJR considers the effect of quantization at the ADCs, so the conditional PMF is characterized in an integral form using the CDF of a normal random variable. Second, Q-BCJR combines multiple observations obtained from receive antennas, so the conditional PMF is characterized in a product form that computes the joint probability of receiving the multiple observations. The relation in \eqref{eq:condPMF_QBCJR} also shows that different pairs of $({\bf s}',{\bf s})\in\bigcup_{k\in\mathcal{K}}\mathcal{V}_{n,k}$ may have the same conditional PMF, so it may not be computed for every pair of $({\bf s}',{\bf s})\in\bigcup_{k\in\mathcal{K}}\mathcal{V}_{n,k}$. This fact contributes to a complexity reduction achieved by Q-BCJR, which will be discussed in the sequel.

After computing the marginal probability using the forward-backward algorithm, the LLR of the $i$-th coded bit is produced by applying \eqref{eq:marginalProb} into \eqref{eq:LLR_eq2}:  
\begin{align}\label{eq:LLR_QBCJR}
	\mathcal{L}_{\rm QBCJR}[i]
	= \log\frac{\sum_{k \in \mathcal{K}_{m_i}(0)} \mathbb{P}({\bf x}[n_i]={\bf x}_k,\tilde{\bf Y})}
	{\sum_{k \in \mathcal{K}_{m_i}(1)} \mathbb{P}({\bf x}[n_i]={\bf x}_k,\tilde{\bf Y})},
\end{align}
for $i\in\{1,\ldots,I_{\rm code}\}$, where $n_i=\lceil\frac{i}{M}\rceil$ and $m_i=i-M(n_i-1)$. In the general case of $\mathcal{W}\neq \emptyset$, the above LLR is the approximation of the true LLR in \eqref{eq:LLR_def} due to the approximate model in \eqref{eq:quantized_dom} and the use of the partial sequence $\tilde{\bf Y}$. In this case, the tightness of the approximation is adjusted by the determination of dominant and weak CIR taps, as discussed in Section III-A. In addition, the approximation becomes tight when the power of weak CIR taps is sufficiently lower than the noise level. The most promising feature of Q-BCJR is that in the case of $\mathcal{W}=\emptyset$, the LLR in \eqref{eq:LLR_QBCJR} becomes the true LLR, so in this extreme case, Q-BCJR is the optimal soft-output detection method for mmWave MIMO systems with low-precision ADCs.

The proposed Q-BCJR is summarized in Algorithm 1. Particularly, normalization steps are added in Step 9 and 13 to prevent ${\alpha}_n({\bf s})$ and ${\beta}_{n-1}({\bf s}')$ from having extremely-low values when $N_{\rm d}$ is large. These additional steps do not affect the resulting LLR in \eqref{eq:LLR_QBCJR}, because any product operation on the marginal probability does not change the LLR values. 
\begin{algorithm}[H]
	\caption{Quantized BCJR (Q-BCJR) algorithm}\label{alg:QBCJR}
	{\small{\begin{algorithmic}[1]
			\STATE Define $\mathcal{S}_n$ for $n\!\in\!\{0,1,\ldots,{N}_{\rm d}\!+\!L_{\mathcal D}\!-\!1\}$ from \eqref{eq:def_state_set}.
			\STATE Define $\mathcal{V}_{n,k}$ for $k\in\mathcal{K}$ and $n\!\in\!\{1,\ldots,{N}_{\rm d}\!+\!L_{\mathcal D}\!-\!1\}$ from \eqref{eq:def_state_trans}.
			\FOR {$n=1$ to ${N}_{\rm d}\!+\!L_{\mathcal D}\!-\!1$}
			\STATE Compute $\gamma_n({\bf s}',{\bf s})$ for $({\bf s}',{\bf s})\in \bigcup_{k\in\mathcal{K}}\mathcal{V}_{n,k}$ from \eqref{eq:gamma} and \eqref{eq:condPMF_QBCJR}.
			\ENDFOR
			\STATE Initialize $\alpha_{0}({\bf s})=\mathbb{I}\{{\bf s}={\bf 0}_{(L_{\mathcal D}-1)N_{\rm tx}}\}$ 
			and $\beta_{{N}_{\rm d}+L_{\mathcal D}-1}({\bf s})=\mathbb{I}\{{\bf s}={\bf 0}_{(L_{\mathcal D}-1)N_{\rm tx}}\}$.
			\FOR {$n=1$ to ${N}_{\rm d}\!-\!1$}
			\STATE	Compute $\alpha_n'({\bf s})=\sum_{{\bf s}'\in\mathcal{S}_{n-1}} \alpha_{n-1}({\bf s}')\gamma_n({\bf s}',{\bf s})$ for ${\bf s}\in\mathcal{S}_n$.
			\STATE	Normalize $\alpha_n({\bf s})=\frac{\alpha_n'({\bf s})}{\sum_{{\bf s}\in\mathcal{S}_n} \alpha_n'({\bf s})}$ for ${\bf s}\in\mathcal{S}_n$.
			\ENDFOR
			\FOR {$n={N}_{\rm d}\!+\!L_{\mathcal D}\!-\!1$ to $2$}
			\STATE	Compute $\beta_{n-1}'({\bf s}')=\sum_{{\bf s}\in\mathcal{S}_n}\beta_n({\bf s})\gamma_n({\bf s}',{\bf s})$ for ${\bf s}'\in\mathcal{S}_{n-1}$.
			\STATE  Normalize $\beta_{n-1}({\bf s}')= \frac{\beta_{n-1}'({\bf s}')}{\sum_{{\bf s}'\in\mathcal{S}_{n-1}} \beta_{n-1}'({\bf s}')}$.
			\ENDFOR
			\FOR {$n=1$ to $N_{\rm d}$}
			\STATE Compute $\mathbb{P}({\bf x}[n]={\bf x}_k,\tilde{\bf Y})$ for $k\in\mathcal{K}$ from \eqref{eq:marginalProb}.
			\ENDFOR 
			\FOR{$i=1$ to $I_{\rm code}$}
			\STATE Compute $\mathcal{L}_{\rm QBCJR}[i]$ from \eqref{eq:LLR_QBCJR} with $n_i=\lceil\frac{i}{M}\rceil$ and $m_i=i-M(n_i-1)$.
			\ENDFOR
	\end{algorithmic}}}
\end{algorithm}

\vspace{-3mm}
From Algorithm~\ref{alg:QBCJR}, we analyze the computational complexity of Q-BCJR when $N_{\rm d}\gg L_{\mathcal D}$. First of all, the complexity order of Steps 3$\sim$5 is $\mathcal{O}\big(\sum_{n=1}^{N_{\rm d}+L_{\mathcal D}-1}|\mathcal{X}|^{|\mathcal{D}_n|}\big)\approx \mathcal{O}(N_{\rm d}|\mathcal{X}|^{|\mathcal{D}_n|})$, because $\mathbb{P}({\bf y}[n]|{\bf s}[n-1]={\bf s}',{\bf s}[n]={\bf s})$ for $({\bf s}',{\bf s})\in \bigcup_{k\in\mathcal{K}}\mathcal{V}_{n,k}$ is determined from $\mathbb{P}({\bf y}[n]|{\bf x}_{\mathcal D}[n])$ for ${\bf x}_{\mathcal D}[n]\in\mathcal{X}^{|\mathcal{D}_n|}$ by the relation in \eqref{eq:condPMF_QBCJR}, where $\mathcal{D}_n =\{\ell: \ell \in \mathcal{D}, 1\leq n-\ell\leq N_{\rm d}\}$ is a subset of $\mathcal{D}$ that only contains the delays of the dominant CIR taps \textit{valid} at time slot $n$. In addition, the complexity order of Steps~7$\sim$10 is $\mathcal{O}\big(\sum_{n=1}^{N_{\rm d}-1}\big|\bigcup_{k\in\mathcal{K}}\mathcal{V}_{n,k}\big|\big)\approx \mathcal{O}(N_{\rm d}|\mathcal{X}|^{L_{\mathcal D}})$, because the product operation in Step 8 is computed for every $\gamma_n({\bf s}',{\bf s})$ for $({\bf s}',{\bf s})\in \bigcup_{k\in\mathcal{K}}\mathcal{V}_{n,k}$, while $\big|\bigcup_{k\in\mathcal{K}}\mathcal{V}_{n,k}\big| = |\mathcal{X}|^{L_{\mathcal D}}$ for $L_{\mathcal D} \leq n \leq N_{\rm d}$ from \eqref{eq:def_state_trans}. Similarly, the complexity order of Steps~11$\sim$14 is $\mathcal{O}\big(\sum_{n=2}^{N_{\rm d}+L_{\mathcal D}-2}\big|\bigcup_{k\in\mathcal{K}}\mathcal{V}_{n,k}\big|\big) \approx \mathcal{O}(N_{\rm d}|\mathcal{X}|^{L_{\mathcal D}})$. Since $|\mathcal{D}_n| \leq L_{\mathcal D}$, the complexity order of Steps~7$\sim$10 and Steps~11$\sim$14 dominates the overall complexity. Therefore, the complexity order of Q-BCJR is given by
\begin{align}\label{eq:complex_QBCJR}
	C_{\rm QBCJR} =  \mathcal{O}\big(N_{\rm d}|\mathcal{X}|^{L_{\mathcal D}}\big) = \mathcal{O}\big(N_{\rm d}2^{ML_{\mathcal D}}\big),
\end{align}
when $N_{\rm d}\gg L_{\mathcal D}$. Recall that $M$ is a symbol-vector modulation level, and $L_{\mathcal D}=\max_{\ell \in\mathcal{D}}\ell+1$ is the \textit{maximum delay index} of the dominant CIR taps. If we do not treat the ISI signals from weak CIR taps as noise (i.e., $\mathcal{W}=\emptyset$), the complexity order of Q-BCJR becomes $\mathcal{O}(N_{\rm d}2^{ML})$. Therefore, our complexity analysis shows that when the maximum delay index of the dominant CIR taps is significantly less than that of the true channel (i.e., $L_{\mathcal D}\ll L$), the proposed Q-BCJR achieves a substantial reduction in the computational complexity provided by the extreme sparsity in the mmWave channels with low-precision ADCs.

%%%%%%%%%%%%%%%%%%%%%%%%%%%%%%%%%%%%%%%%%%%%%%%%%%%%%%%%%%%%%%%%%%%%%%%%%%%%%%%%%%%%%%%%%%%%
%%%%%%%%%%%%%%%%%%%%%%%%%%%%%%%%%%%%%%%%%%%%%%%%%%%%%%%%%%%%%%%%%%%%%%%%%%%%%%%%%%%%%%%%%%%%
\subsection{Quantized Belief-Propagation (Q-BP)}
One drawback of Q-BJCR is that its computational complexity is not affordable in practical systems when the maximum delay of the dominant CIR taps is large (i.e., $L_{\mathcal D}\gg1$). To overcome this limitation, we also develop a low-complexity soft-output detection method called \textit{quantized belief-propagation (Q-BP)} which requires a significantly lower complexity than Q-BCJR does. 

\begin{figure*}[t]
	\centering
	\subfigure[Original factor graph]
	{\epsfig{file=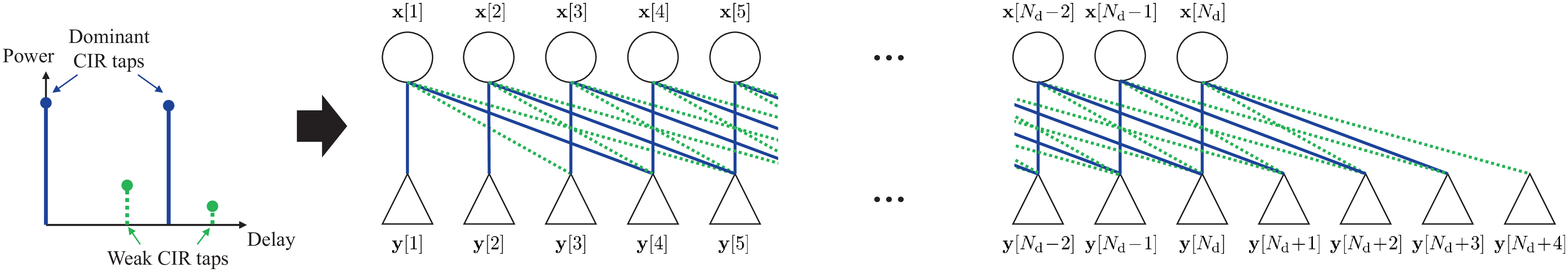, width=16cm}}
	\hfill
	\centering
	\subfigure[Sparse factor graph]
	{\epsfig{file=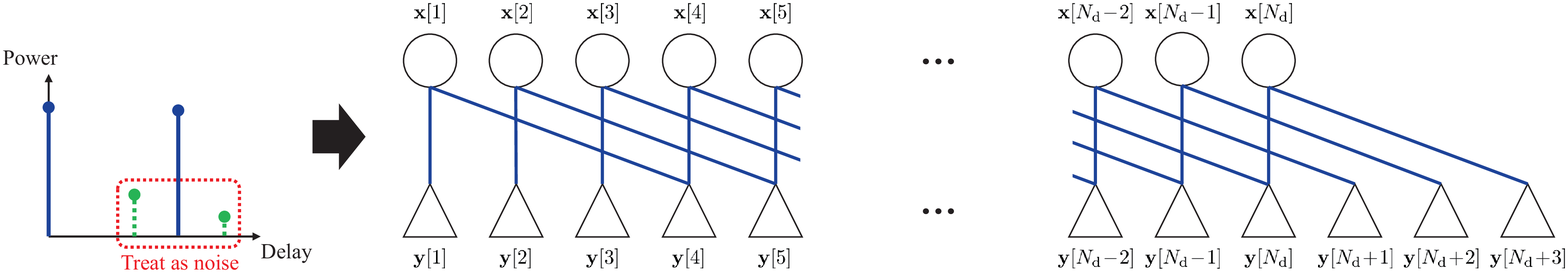, width=16cm}}
	\caption{The comparison between (a) the original factor graph and (b) the sparse factor graph constructed when the delay sets of dominant and weak CIR taps are given by $\mathcal{D}=\{0,3\}$ and $\mathcal{W}=\{2,4\}$, respectively.} \vspace{-3mm} \label{fig:FG}
\end{figure*}

The key idea of Q-BP is to create a sparse factor graph based on the extremely sparse ISI channel model in Section III. Then it computes the LLRs in an iterative fashion by using the BP algorithm based on this sparse factor graph. To present this idea, we first explain how to create such the sparse factor graph when dominant and weak CIR taps are given. An original factor graph that describes the input-output relation of the transmitted symbol vectors and the quantized received vectors consists of $N_{\rm d}$ variable nodes and $N_{\rm d}+L-1$ check nodes. Each variable node is associating with the transmitted symbol vector at each time slot, and each check node is associating with the quantized received vector at each time slot. A check node is connected with a variable node by an \textit{edge}, if the quantized vector associating with the check node depends on the symbol vector associating with the variable node by a nonzero CIR tap. For example, in Fig.~\ref{fig:FG}(a), the original factor graph of a mmWave MIMO system when $\mathcal{D}=\{0,3\}$ and $\mathcal{W}=\{2,4\}$ is illustrated, where circles are the variable nodes, triangles are the check nodes, and solid and dotted lines are the edges associating with the dominant CIR taps and the weak CIR taps, respectively. The sparse factor graph of Q-BP is constructed by ignoring the edges associating with the weak CIR taps (dotted lines) among all the edges of the original factor graph, as illustrated in Fig.~\ref{fig:FG}(b). Thanks to the sparsity in the mmWave channel, the constructed sparse factor graph has a less number of edges than the original factor graph does. It is also noticeable that the number of \textit{valid} check nodes  in the sparse factor graph is $N_{\rm d}+L_{\mathcal D}-\min_{\ell \in\mathcal{D}}\ell - 1$, which is less than that of the original factor graph.

By harnessing the sparse factor graph, Q-BP computes the APP $\mathbb{P}({\bf x}[n]={\bf x}_k|\tilde{\bf Y})$ in an iterative fashion, in which variable nodes and check nodes iteratively exchange their local beliefs, called \textit{messages}, through the edges of the sparse factor graph. We explain how to determine the messages from the variable nodes and the check nodes with details.

\subsubsection{Message from variable node to check node}
The $n$-th variable node sends $|\mathcal{K}|$ messages that contain the extrinsic information of the conditional PMF for the corresponding symbol vector ${\bf x}[n]$. These messages are passed to the $(n+\ell)$-th check node for $\ell\in\mathcal{D}$. By assuming that the transmission of each possible symbol vector is equally likely, the $k$-th message from the $n$-th variable node to the $(n+\ell)$-th check node, namely $T_n^{n+\ell}(k)$, is determined as \cite{Pearl:98}:
\begin{align}\label{eq:T_def}
	T_{n}^{n+\ell}(k) = \frac{\prod_{m\in\mathcal{D}\setminus\{\ell\}} R_{n+m}^n(k) }{\sum_{j\in\mathcal{K}} \prod_{m\in\mathcal{D}\setminus\{\ell\}} R_{n+m}^n(j) },
\end{align}
for $\ell \in\mathcal{D}$ and $k\in\mathcal{K}$, where $R_{n+m}^n(j)$ is the $j$-th message from the $(n+m)$-th check node. The message of $T_{n}^{n+\ell}(k)$ propagates the marginal probability of the event $\{{\bf x}[n]={\bf x}_k\}$ using the quantized observations except the quantized received signal at time slot $n+\ell$. At the initial stage of the algorithm, no information is available at each variable node; thereby, all messages from the variable nodes are initialized as $T_{n}^{n+\ell}(k)=\frac{1}{K}$ for $\ell \in\mathcal{D}$ and $k\in\mathcal{K}$.

\subsubsection{Message from check node to variable node}
The $n$-th check node sends $|\mathcal{K}|$ messages that contain the a-posteriori information of the $(n-\ell)$-th transmitted symbol vector obtained from the quantized received signal at time slot $n$. These messages are passed to the $(n-\ell)$-th variable node for $\ell\in\mathcal{D}_n$, where $\mathcal{D}_n =\{\ell: \ell \in \mathcal{D}, 1\leq n-\ell\leq N_{\rm d}\}$ is a subset of $\mathcal{D}$ that only contains the delays of the dominant CIR taps \textit{valid} at time slot $n$. By assuming that all the incoming messages from the connected variable nodes are independent, the $k$-th message from the $n$-th check node to the $(n-\ell)$-th variable node, namely $R_n^{n-\ell}(k)$, is determined as \cite{Pearl:98}:
\begin{align}\label{eq:R_def}
	R_{n}^{n-\ell}(k) 
	=  \sum_{k_m\in\mathcal{K},m\in\mathcal{D}_n\setminus\{\ell\}} &\mathbb{P}\left( {\bf y}[n] \Big| {\bf x}[n\!-\!\ell]\!=\!{\bf x}_k,\big\{{\bf x}[n\!-\!m]\!=\!{\bf x}_{k_m}\big\}_{m\in\mathcal{D}_n\setminus\{\ell\}}\right) \prod_{m\in\mathcal{D}_n\setminus\{\ell\}} T_{n-m}^n(k_m),
\end{align}
for $\ell\in\mathcal{D}_n$ and $k\in\mathcal{K}$. The conditional PMF term in \eqref{eq:R_def} is computed from \eqref{eq:condPMF_dom} by applying ${\bf x}_{\mathcal D}[n]$ that associates with the events $\{{\bf x}[n-\ell]={\bf x}_k\}$ and $\{{\bf x}[n-m]={\bf x}_{k_m}\}_{m\in\mathcal{D}_n\setminus\{\ell\}}$. The above message propagates the APP of the event $\{{\bf x}[n]={\bf x}_k\}$ based on the incoming messages and its own observation ${\bf y}[n]$. As can be seen in \eqref{eq:R_def}, when determining the message of $R_{n}^{n-\ell}(k)$, the incoming messages from the connected variable nodes are utilized except the one from the $(n-\ell)$-th variable node.

After iteratively exchanging the messages between the check node and the variable node, each variable node is assumed to obtain the marginal distribution of the transmitted symbol vector that is sufficiently learned for the given quantized observations. Then the LLR of the $i$-th coded bit is obtained as
\begin{align}\label{eq:LLR_QBP}
	\mathcal{L}_{\rm QBP}[i] 
	= \log\frac{\sum_{k \in \mathcal{K}_{m_i}(0)} \prod_{\ell\in\mathcal{D}} R_{n_i+\ell}^{n_i}(k) }{\sum_{k \in \mathcal{K}_{m_i}(1)} \prod_{\ell\in\mathcal{D}} R_{n_i+\ell}^{n_i}(k) },
\end{align}
for $i\in\{1,\ldots,I_{\rm code}\}$. One major drawback of Q-BP is that when the sparse factor graph is not cycle free, the convergence of the algorithm is not guaranteed, so it may fail to provide the true LLRs. This drawback, however, does not have a significant impact on the detection performance as discussed in \cite{Colavolpe:05,Kaynak:05}. A simple intuition is that in most channel realizations, there exists an edge in the cycle that associates with a CIR tap having a relatively small power than others. Such edge effectively \textit{cuts} the cycle, so the effect of the cycle becomes negligible.

The proposed Q-BP is summarized in Algorithm~\ref{alg:QBP}, where $N_{\rm it}$ is the number of iterations that determines the performance-complexity tradeoff achieved by Q-BP. Since the structure of the factor graph may significantly vary according to channel realizations, we simply adopt a \textit{flooding (parallel) schedule} as in \cite{Kurkoski:02,Colavolpe:05,Kaynak:05} which does not depend on the factor graph structure. 
\begin{algorithm}[H]
	\caption{Quantized Belief Propagation (Q-BP) algorithm}\label{alg:QBP}
	{\small{\begin{algorithmic}[1]
			\STATE Initialize $T_{n}^{n+\ell}(k)=\frac{1}{|\mathcal{K}|}$ for $k\in\mathcal{K}$, $\ell\in\mathcal{D}$, and $n\in\{1,\ldots,N_{\rm d}\}$.
			\FOR {$it=1$ to $N_{\rm it}$}
				\FOR {$n=1$ to ${N}_{\rm d}+L_{\mathcal D}-1$}
					\STATE Compute $R_{n}^{n-\ell}(k)$ for $k\in\mathcal{K}$  and $\ell\in\mathcal{D}_n$ from \eqref{eq:R_def} and \eqref{eq:condPMF_dom}.
				\ENDFOR
				\FOR {$n=1$ to $N_{\rm d}$}
					\STATE	Compute $T_{n}^{n+\ell}(k)$ for $k\in\mathcal{K}$ and $\ell\in\mathcal{D}$ from \eqref{eq:T_def}.
				\ENDFOR
			\ENDFOR
			\FOR{$i=1$ to $I_{\rm code}$}
				\STATE Compute $\mathcal{L}_{\rm QBP}[i]$ from \eqref{eq:LLR_QBP} with $n_i=\lceil\frac{i}{M}\rceil$ and $m_i=i-M(n_i-1)$.
			\ENDFOR
	\end{algorithmic}}}
\end{algorithm}

\vspace{-3mm}
From Algorithm~\ref{alg:QBP}, we analyze the computational complexity of Q-BP when $N_{\rm d}\gg L_{\mathcal D}$. First of all, the complexity order of Steps 3$\sim$5 is 
\begin{align}\label{eq:complex_QBP_3_5}
	\mathcal{O}\left(N_{\rm it}\sum_{n=1}^{{N}_{\rm d}+L_{\mathcal D}-1} \sum_{k \in \mathcal{K}}|\mathcal{D}_n||\mathcal{X}|^{|\mathcal{D}_n|-1}\right) 
	%= \mathcal{O}\left(N_{\rm it}\sum_{n=1}^{\bar{N}_{\rm d}} |\mathcal{D}_n|\times 2^{M|\mathcal{D}_n|}\right)
	\overset{(a)}{\approx} \mathcal{O}\big(N_{\rm it}N_{\rm d}|\mathcal{D}||\mathcal{X}|^{|\mathcal{D}|}\big),
\end{align}
because $R_{n}^{n-\ell}(k)$ determined from \eqref{eq:R_def} requires $|\mathcal{X}|^{|\mathcal{D}_n|-1}$ computations of the conditional PMF for $n\in\{1,\ldots,{N}_{\rm d}+L_{\mathcal D}-1\}$, $\ell\in\mathcal{D}_n$,  $k\in\mathcal{K}$, and each iteration. The approximation of (a) in \eqref{eq:complex_QBP_3_5} holds because $\mathcal{D}_n=\mathcal{D}$ for most cases in $1\leq n \leq N_{\rm d}$ with ${N}_{\rm d}+L_{\mathcal D}-1\approx N_{\rm d}$ when  $N_{\rm d}\gg L_{\mathcal D}$. Since the complexity order of Steps 3$\sim$5 clearly dominates the overall complexity, the complexity order of Q-BP is given by 
\begin{align}\label{eq:complex_QBP}
	C_{\rm QBP} = \mathcal{O}\big(N_{\rm it}N_{\rm d}|\mathcal{D}||\mathcal{X}|^{|\mathcal{D}|}\big) = \mathcal{O}\big(N_{\rm it}N_{\rm d}|\mathcal{D}|2^{M|\mathcal{D}|}\big),
\end{align}
when $N_{\rm d}\gg L_{\mathcal D}$. The comparison between \eqref{eq:complex_QBCJR} and \eqref{eq:complex_QBP} shows that the complexity order of Q-BP is only ${N_{\rm it}|\mathcal{D}|}{2^{-M(L_{\mathcal D}-|\mathcal{D}|)}}$ of that of Q-BCJR. Since in most channel realizations, the \textit{number} of the dominant CIR taps is smaller than the \textit{maximum delay index} of them, Q-BP achieves a significant reduction in the computational complexity compared to Q-BCJR. 
%Another advantage of Q-BP is that its applicability is not limited by channel environments, since the computational complexity can be adjusted by the design parameters $|\mathcal{D}|$ and $N_{\rm it}$. %The effect of these parameters to the detection performance will be explored in Section IV by simulations.

\vspace{1mm}
{\bf Remark (The effect of channel sparsity):}
The performance-complexity tradeoff achieved by the proposed detection methods (Q-BP and QBCJR) improves as the delay-domain sparsity level in mmWave channels increases, because the computational complexity of both methods reduces with the channel sparsity level as discussed in Section IV. For this reason, the proposed methods are effective solutions not only in mmWave channels, but also in other high-frequency channels or line-of-sight (LOS) channels that have a strong sparsity level. It is also noticeable that even for non-sparse channels, the proposed methods can maintain a fair level of the FER performance at the cost of the computational complexity.

%%%%%%%%%%%%%%%%%%%%%%%%%%%%%%%%%%%%%%%%%%%%%%%%%%%%%%%%%%%%%%%%%%%%%%%%%%%%%%%%%%%%%%%%%%%%
%%%%%%%%%%%%%%%%%%%%%%%%%%%%%%%%%%%%%%%%%%%%%%%%%%%%%%%%%%%%%%%%%%%%%%%%%%%%%%%%%%%%%%%%%%%%
%%%%%%%%%%%%%%%%%%%%%%%%%%%%%%%%%%%%%%%%%%%%%%%%%%%%%%%%%%%%%%%%%%%%%%%%%%%%%%%%%%%%%%%%%%%%
\section{Simulation Results}\label{sec:simul}
%In this section, using simulations, we evaluate the detection performance of the proposed soft-output detection methods for mmWave MIMO systems with low-precision ADCs, compared to the existing OFDM-based detection methods. 
In this section, using simulations, we evaluate the performance of the soft-output detection methods proposed in Section IV for mmWave MIMO systems with low-precision ADCs. We also evaluate the performance-complexity tradeoff achieved by the dominant-tap-selection algorithm proposed in Section III-B.

\subsection{Simulation Setting}
In simulations, we adopt a $B$-bit uniform scalar quantizer in which the set of quantization alphabets is set to be $\mathcal{Q}=\{-1,+1\}$ and $\mathcal{Q}=\{-1.125,-0.375,0.375,1.125\}$ for $B=1$ and $B=2$, respectively, such that $b_p=\frac{q_{p}+q_{p+1}}{2}$ for $p\in\{1,\ldots,2^B-1\}$. 
%we adopt a $B$-bit uniform scalar quantizer with a step size $\Delta_{B}$, in which the $p$-th quantization alphabet is $q_p=\frac{\Delta_{B}}{2}(-2^{B}-1+2p)$ for $p\in\{1,\ldots,2^B\}$ and the $p$-th quantization bin boundary is $b_p=\frac{\Delta_{B}}{2}(-2^{B}+2p)$ for $p\in\{1,\ldots,2^B-1\}$ with $b_0=-\infty$ and $b_{2^B}=\infty$. Particularly, we set the step size of the quantizer as $\Delta_{B}=\{2, 0.75\}$ for $B=\{1,2\}$, respectively. 
For a channel code, we adopt a 1/2-rate LDPC code with $I_{\rm info}=336$ and $I_{\rm code}=672$ from the IEEE 802.11ad standardization \cite{IEEE}, along with a soft-input belief-propagation channel decoder \cite{Richardson}. As a control variable, we use the average signal-to-noise ratio (SNR) per bit defined as 
\begin{align}\label{eq:EbN0}
	\frac{E_b}{N_0}=\frac{\mathbb{E}\big[\|{\bf x}[n]\|^2\big]}{\sigma^2\log_2|\mathcal{X}|}= \frac{N_{\rm tx}}{M\sigma^2}.
\end{align}
For the proposed dominant-tap-selection algorithm, we set the NMSE threshold as $\epsilon_{\rm th}=0.1$ unless otherwise specified. For the proposed Q-BP, we set the number of iterations as $N_{\rm it}=3$ which is numerically shown to be a value that makes Algorithm 3 converged for most system settings.Since the optimal design of the transmit digital BF is still an open problem for the mmWave MIMO systems with low-precision ADCs \cite{Mo:BF:17}, we assume the trivial digital BF at the transmitter (i.e., ${\bf F}_{\rm tx}^{\rm BB}={\bf I}_{N_{\rm tx}}$) to avoid undesirable FER degradation caused by the use of a suboptimal BF.

For imperfect CSIR case, we adopt a least-squares (LS) channel estimation method with $T_{\rm p}$ pilot signals to estimate the CIRs of the effective channel in \eqref{eq:equi_ch}. Let ${\bf X}_{\rm p}=\big[{\bf x}_{\rm p}[1],\ldots,{\bf x}_{\rm p}[T_{\rm p}]\big] \in\mathbb{C}^{N_{\rm tx}\times T_{\rm p}}$ be a pilot signal matrix, where ${\bf x}_{\rm p}[n] \in \mathbb{C}^{N_{\rm tx}}$ is the $n$-th pilot signal vector, and $T_{\rm p}$ is the length of the pilot signals such that $T_{\rm p}\geq L(N_{\rm tx}\!-\!1)\!+\!1$. The pilot signals are randomly chosen to satisfy the orthogonality condition of ${\bf X}_{\rm p}{\bf X}_{\rm p}^{\sf H}=\sqrt{T_{\rm p}}{\bf I}_{N_{\rm tx}}$. According to the signal model in Section II-B, the unquantized received signal at the $i$-th receive RF chain during the pilot transmission is expressed as 
\begin{align}\label{eq:CE_received}
	{\bf r}_{\rm p}^{(i)}  = \bar{\bf X}_{\rm p} {\bf h}_{i} + {\bf z}_{\rm p}^{(i)},
\end{align}
for $i\in\{1,\ldots,N_{\rm rx}\}$, where ${\bf h}_{i}^\top$ is the $i$-th row of the CIR matrix $\big[{\bf H}[0],\cdots,{\bf H}[L-1]\big]$, $\bar{\bf X}_{\rm p} \in \mathbb{C}^{(T_{\rm p}+L-1)\times LN_{\rm tx}}$ is a toeplitz-type matrix that consists of the pilot signals, and ${\bf z}_{\rm p}^{(i)}\in \mathbb{C}^{T_{\rm p}+L-1}$ is the noise vector at the $i$-th receive RF chain. Then, by applying the LS estimation method to the quantized signal, the estimate for ${\bf h}_{i}$ is obtained as
\begin{align}\label{eq:CE_estimated}
	\hat{\bf h}_{i} =  (\bar{\bf X}_{\rm p}^{\sf H}\bar{\bf X}_{\rm p})^{-1}\bar{\bf X}_{\rm p}^{\sf H}{\bf y}_{\rm p}^{(r)}
\end{align}	
where ${\bf y}_{\rm p}^{(i)} = Q({\rm Re}\{{\bf r}_{\rm p}^{(i)}\})+jQ({\rm Im}\{{\bf r}_{\rm p}^{(i)}\})$. Consequently, the estimate of the CIR matrix is given by $\big[\hat{\bf h}_{1},\cdots,\hat{\bf h}_{N_{\rm rx}}\big]^\top$.

For channel generation, we consider three different channel models described below. 
\begin{itemize}
	\item {\em 6-tap Exp-PDP channel:} In this model, the CIR taps of the mmWave channels are modeled by independent Rayleigh fading CIRs that follow a $6$-tap exponentially-decaying power-delay profile with an exponent 1. 
	\item {\em 28-GHz and 72-GHz NLOS channels:} In these models, 28-GHz and 73-GHz NLOS channels are implemented according to the measurement-based model in \cite{Samimi:16}. Parameters for the implementation are the same as those for the numerical example in Section II-A. The channel length $L$ is chosen to be less than 336 to ensure that the channel length is smaller than the data block length.  
\end{itemize}

For performance comparison, we consider one optimal detection method and three existing OFDM-based detection methods described below.
\begin{itemize}
	\item {\em BCJR:} This method is the optimal detection method for conventional mmWave MIMO systems with infinite-precision ADCs, which applies the original BCJR algorithm to compute the exact LLRs in the time domain. 
	
	\item {\em OFDM-Convex:} This method performs joint-subcarrier data equalization by solving a convex optimization problem using the FASTA algorithm proposed in \cite{Studer:16}.
	
	\item {\em OFDM-Bussgang:} This method performs per-subcarrier data equalization by linearizing the quantized received signal based on Bussgang's theorem \cite{Bussgang} under the assumption of the Gaussian signaling.
	
	\item {\em OFDM-MMSE:} This method performs per-subcarrier data equalization by ignoring the quantization effect at the ADCs (i.e., by assuming ${\bf y}[n]={\bf r}[n]$). 
\end{itemize}
Particularly for the OFDM-based methods, we set the length of cyclic prefix (CP) as $L-1$, and use the normalized noise power $\frac{N_{\rm d}+L-1}{N_{\rm d}}\sigma^2$ to reflect the power consumed by the CP.

%The quantization alphabet $\mathcal{Y}$ and the quantization bin boundaries $\{b_k\}$ are set using the maximum-output entropy (MOE) quantizer in \cite{Messerschmitt:71}. 

%%%%%%%%%%%%%%%%%%%%%%%%%%%%%%%%%%%%%%%%%%%%%%%%%%%%%%%%%%%%%%%%%%%%%%%%%%%%%%%%%%%%%%%%%%%%
%%%%%%%%%%%%%%%%%%%%%%%%%%%%%%%%%%%%%%%%%%%%%%%%%%%%%%%%%%%%%%%%%%%%%%%%%%%%%%%%%%%%%%%%%%%%
\subsection{FER Performance}

\begin{figure*}[t]
	\centering
	\subfigure[Perfect CSIR]
	{\epsfig{file=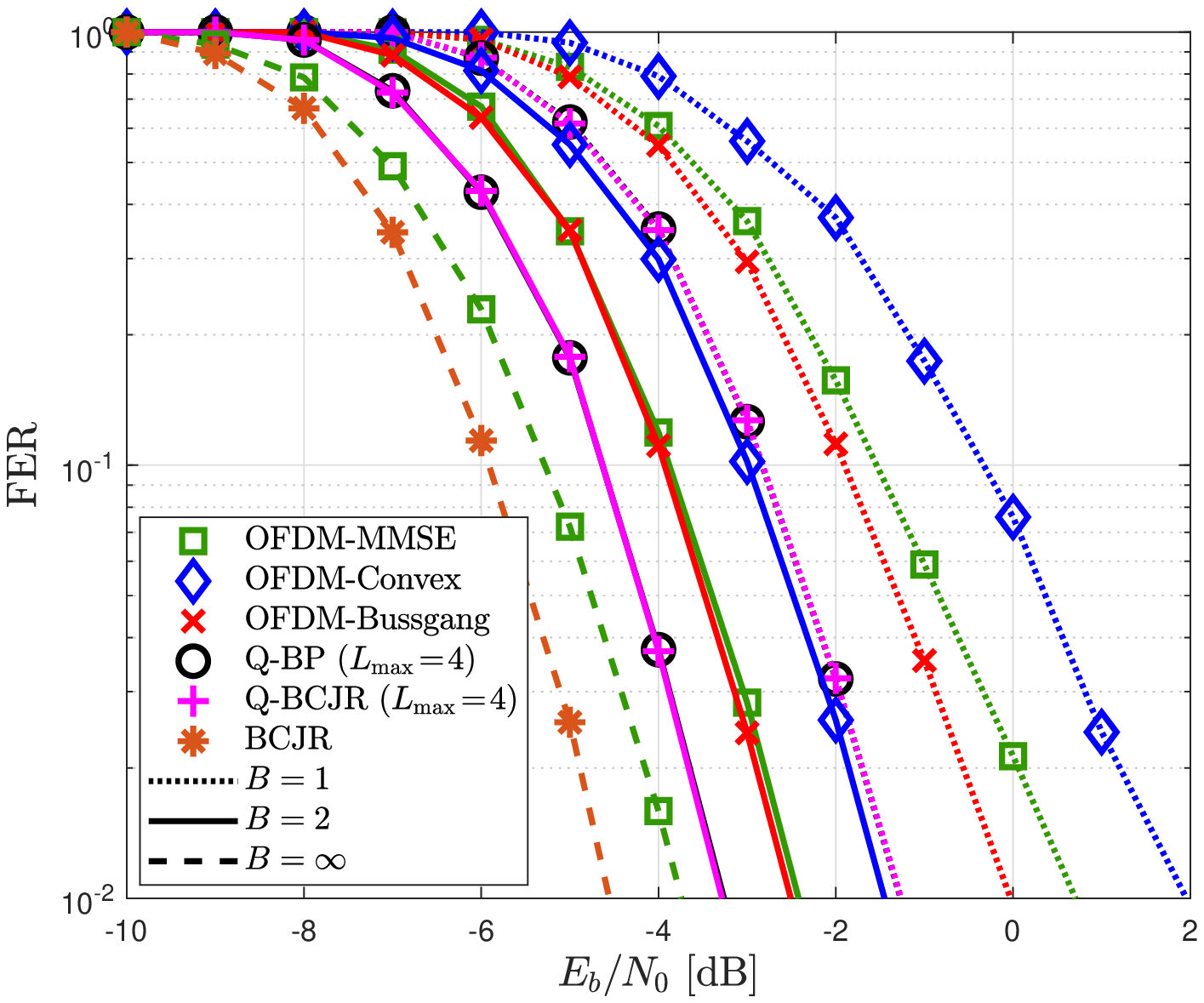, width=7cm}}
	\subfigure[Imperfect CSIR ($T_{\rm p}=24$)]
	{\epsfig{file=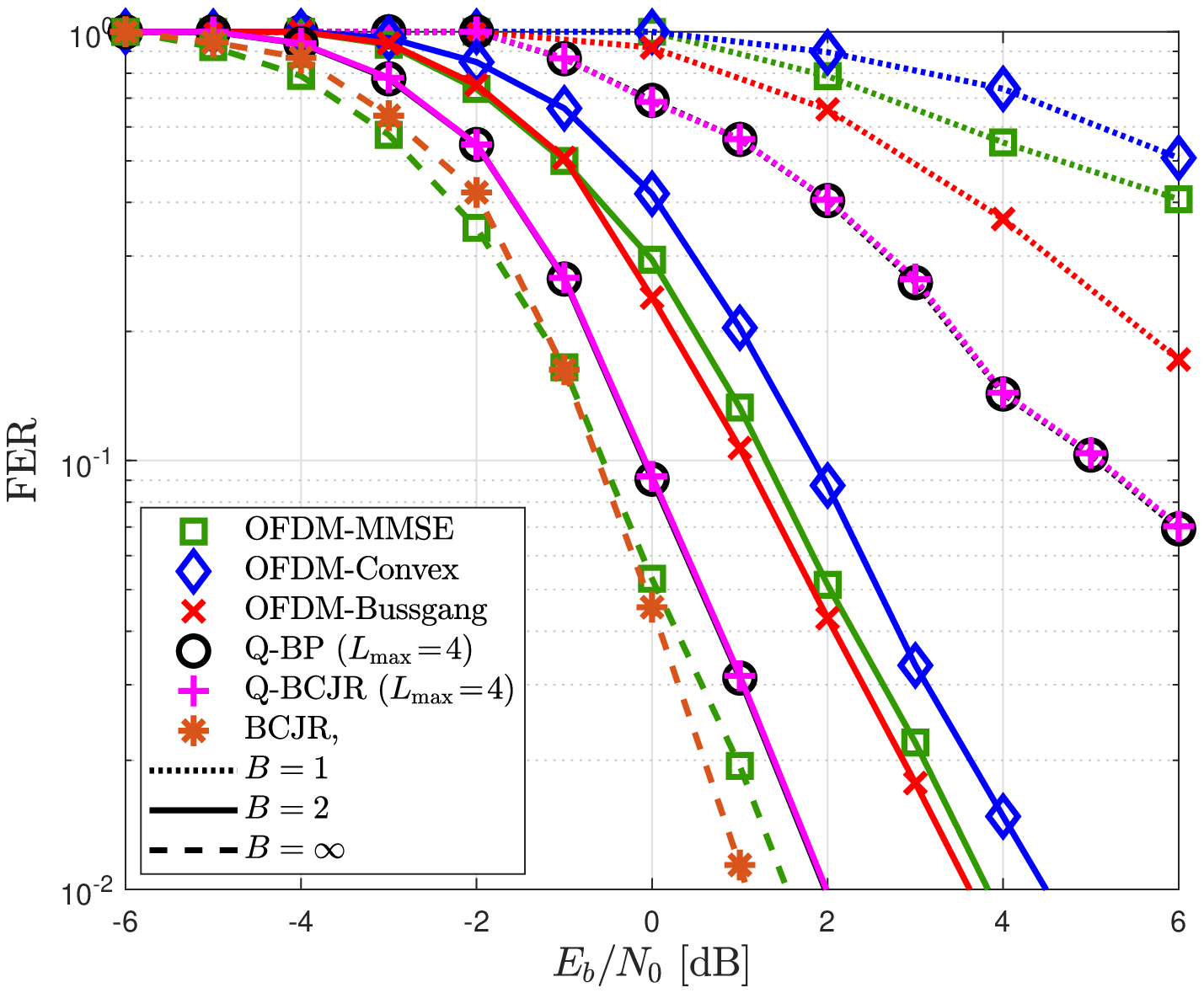, width=7cm}}
	\vspace{-2mm}\caption{The FER vs. $E_b/N_0$ of the proposed Q-BCJR, the proposed Q-BP, and the existing soft-output detection methods under the 6-tap Exp-PDP channel model with $N_{\rm tx}=2$, $N_{\rm rx}=4$, and BPSK.}\vspace{-3mm}
	\label{fig:2x4}
\end{figure*}
Fig.~\ref{fig:2x4} compares the FER performances of the proposed and the existing soft-output detection methods under the 6-tap Exp-PDP channel model with $N_{\rm tx}=2$, $N_{\rm rx}=4$, and binary phase shift keying (BPSK). Fig.~\ref{fig:2x4}(a) shows that when perfect CSIR is available, the proposed methods with 2-bit ADCs perform very close to the optimal performance achieved by the BCJR algorithm. This result demonstrates that the use of 2-bit ADCs may not cause a significant FER loss compared to infinite-bit ADC case, provided that a proper detection method (e.g., Q-BCJR or Q-BP) is employed at the receiver. The proposed methods also outperform the existing OFDM-based methods for both one-bit and two-bit ADC cases. Fig.~\ref{fig:2x4}(b) shows that when CSIR is imperfect, the performance gap between the proposed methods and the BCJR algorithm is further reduced, while the performance gain over the existing OFDM-based methods becomes larger. The reason for this result is that when treating the weak CIR taps as the additional noise, this noise also acts like a compensation term for a channel estimation error; thereby, the proposed methods are more robust to the channel estimation error compared to other methods. Among the existing OFDM-based methods, OFDM-Bussgang shows the best FER performance as it properly considers the effect of the quantization noise by using Bussgang's theorem. Although OFDM-Convex performs the joint-subcarrier soft-output detection, it still suffers from the severe FER degradation due to the lack of the post-equalization signal-to-interference-plus-noise ratio information when computing the LLRs, as reported in \cite{Studer:16}.

\begin{figure*}[t]
	\centering
	\subfigure[BPSK, 28-GHz NLOS, and perfect CSIR]
	{\epsfig{file=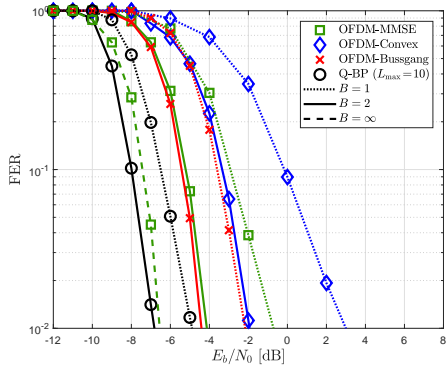, width=6.7cm}}
	\subfigure[4-QAM, 28-GHz NLOS, and perfect CSIR]
	{\epsfig{file=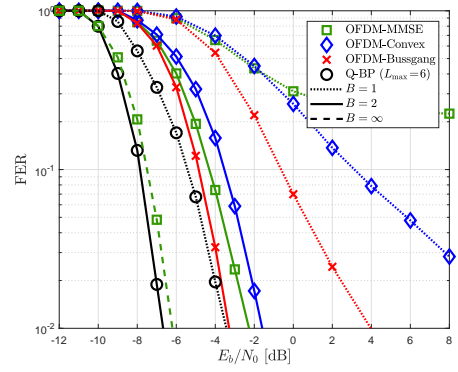, width=6.7cm}}
	\subfigure[BPSK, 73-GHz NLOS, and perfect CSIR]
	{\epsfig{file=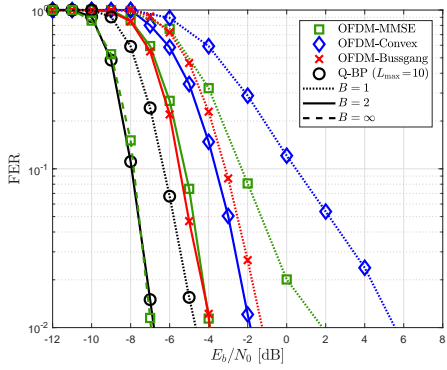, width=6.7cm}}
	\subfigure[BPSK, 28-GHz NLOS, and imperfect CSIR ($T_{\rm p}=2L$)]
	{\epsfig{file=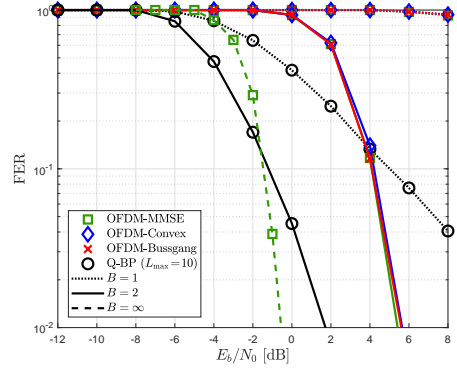, width=6.7cm}}
	\vspace{-2mm}
	\caption{The FER vs. $E_b/N_0$ of the proposed Q-BP, and the existing OFDM-based detection methods under the 28-GHz and the 73-GHz NLOS channel models with $N_{\rm tx}=1$ and $N_{\rm rx}=8$.}\vspace{-3mm}
	\label{fig:1x8}
\end{figure*}
Fig.~\ref{fig:1x8} compares the FER performances of the proposed and the existing soft-output detection methods\footnote{In this simulation, the performances of Q-BCJR and the original BCJR algorithm are not presented because their complexities are not affordable in the considered channel model whose maximum delay index can be high.} under the 28-GHz and the 73-GHz NLOS channel models with $N_{\rm tx}=1$ and $N_{\rm rx}=8$. When adopting 4-quadrature-amplitude-modulation (4-QAM), the results are averaged only over the scenarios that the proposed Q-BP requires an affordable level of the computational complexity; thereby, in this case, the channels that have a less than 6 dominant CIR taps are simulated. Figs.~\ref{fig:1x8}(a) and (b) show that when employing one- or two-bit ADCs under the 28-GHz NLOS channel, the proposed Q-BP outperforms the existing OFDM-based methods regardless of the modulation set. Since the  fast fading characteristics of the 28-GHz and the 73-GHz NLOS channels are not significantly different as shown in Fig. 2(a), the results in Fig. \ref{fig:1x8}(c) are similar to those in Fig.~\ref{fig:1x8}(a). The comparison between Figs.~\ref{fig:1x8}(a) and (d) reveals that the FER gain achieved by the proposed Q-BP becomes larger for the imperfect CSIR case than the perfect CSIR case. This larger gain is obtained by the robustness of the Q-BP to the channel estimation error as already discussed in Fig. \ref{fig:2x4}. One noticeable observation is that the proposed Q-BP with 2-bit ADCs even outperforms OFDM-MMSE with infinite-resolution ADCs in low SNR regime, by computing near-optimal LLR values at the expense of the computational complexity. This gain, however, vanishes as SNR increases, since Q-BP cannot overcome a fundamental diversity loss caused by the use of low-precision ADCs. Another interesting observation is that the performance gain of the proposed Q-BP over the existing methods becomes larger for the realistic mmWave channel model than that for the short-delay channel model in Fig. \ref{fig:2x4}. The reason for this result is that when employing low-precision ADCs, the larger the number of the CIR taps, the larger the inter-subcarrier interference that degrades the performance of the frequency-domain equalization.

%%%%%%%%%%%%%%%%%%%%%%%%%%%%%%%%%%%%%%%%%%%%%%%%%%%%%%%%%%%%%%%%%%%%%%%%%%%%%%%%%%%%%%%%%%%%
%%%%%%%%%%%%%%%%%%%%%%%%%%%%%%%%%%%%%%%%%%%%%%%%%%%%%%%%%%%%%%%%%%%%%%%%%%%%%%%%%%%%%%%%%%%%
\subsection{Performance-Complexity Tradeoff}
\begin{figure}
	\centering
	\includegraphics[width=7cm]{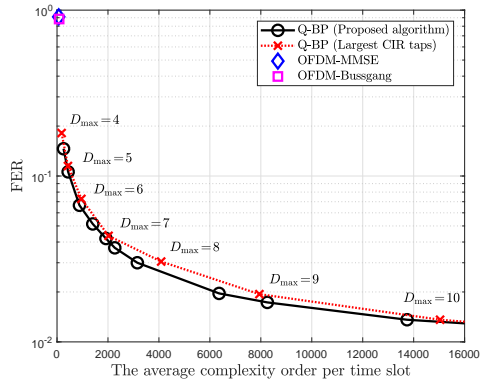} \vspace{-0.3cm}\caption{The FER vs. the average complexity order per time slot of the proposed Q-BP with different dominant-tap-selection algorithms and the existing OFDM-based methods under the 28-GHz NLOS channel model with $N_{\rm tx}=1$, $N_{\rm rx}=8$, $E_b/N_0=-7$ dB, and 2-bit ADCs.} \vspace{-3mm} \label{fig:Tradeoff}
\end{figure}

Fig.~\ref{fig:Tradeoff} compares the performance-complexity tradeoff achieved by the proposed Q-BP and the existing OFDM-based methods (OFDM-MMSE and OFDM-Bussgang) under the 28-GHz NLOS channel model with $N_{\rm tx}=1$, $N_{\rm rx}=8$, BPSK, $E_b/N_0=-7$ dB, and 2-bit ADCs. For the Q-BP, we also compare the tradeoff performances of two different  dominant-tap-selection algorithms: 1) the proposed algorithm (Algorithm 1), and 2) a simple algorithm that selects the $\min(D_{\rm max},|\mathcal{S}|)$ largest CIR taps with respect to the channel power $\|{\bf H}[\ell]\|_{\rm F}^2$. We consider the FER and the average computational complexity order per time slot\footnote{This order is given by $N_{\rm it}|\mathcal{D}|2^{M|\mathcal{D}|}$, $N_{\rm rx}\log_2N_{\rm d}$, and $N_{\rm rx}(\log_2 N_{\rm d}+2)$ for the proposed Q-BP, OFDM-MMSE, and OFDM-Bussgang, respectively.} to evaluate the performance and the complexity, respectively. For the proposed dominant-tap-selection algorithm, we numerically choose the number of the maximum dominant CIR taps, $D_{\rm max}$, and the NMSE threshold, $\epsilon_{\rm th}$, that maximize the tradeoff.

Fig~\ref{fig:Tradeoff} shows that the proposed Q-BP achieves a significant FER reduction compared to the existing OFDM-based methods, by increasing the computational complexity. It is also shown that the performance-complexity tradeoff achieved by the Q-BP is adjusted by the parameters of the dominant-tap-selection algorithm. Among two different selection algorithms, the proposed algorithm provides a better performance-complexity tradeoff than the algorithm that simply selects the largest CIR taps. This additional tradeoff gain is not significant because the NMSE criterion of the proposed algorithm is also minimized by selecting the largest CIR taps when the power difference among the CIR taps is large. Nevertheless, the proposed algorithm is still useful to improve the tradeoff achieved by the proposed Q-BP, as the NMSE criterion effectively reduces the modeling error of the Q-BP even when the CIR taps have a similar power. It is also noticeable that this gain vanishes as the complexity order increases, because both dominant-tap-selection algorithms may select all nonzero CIR taps when $D_{\rm max}$ is sufficiently large.

\vspace{-0.1cm}
\section{Conclusion}\vspace{-0.02cm}
In this paper, we have studied a soft-output detection problem in mmWave MIMO systems with low-precision ADCs. Our key strategy is to construct the extremely sparse ISI channel model by jointly exploiting the delay-domain sparsity in the mmWave channel and a high quantization noise caused by low-precision ADCs. Based on this channel model, we have developed two detection methods, referred to as Q-BCJR and Q-BP, by applying the forward-and-backward algorithm and the BP algorithm, respectively. In particular, Q-BCJR has been shown to provide the near-optimal LLR values, while Q-BP achieves a significant reduction in the computational complexity compared to Q-BCJR. Simulation results have shown that when employing one- or two-bit ADCs, both Q-BCJR and Q-BP provide significant FER gains compared to the existing OFDM-based detection methods. 
%Therefore, by simulations, we have demonstrated that the time-domain equalization using the sparsity is a promising solution for the mmWave MIMO systems with low-precision ADCs, in order to overcome a significant performance loss caused by the use of low-precision ADCs.

An important direction for future research is to develop a robust soft-output detection method that overcomes the effect of a channel estimation error at the receiver. Another important extension is to study the joint design of the soft-output detection method and the channel decoder by exploiting both the delay-domain sparsity of the mmWave channel and the structure of the code construction. This extension would further improve the performance of the mmWave systems with low-precision ADCs.

\appendices

\section{Proof of Proposition~1}\label{sec:Apdx1}
This proof is a simple extension of the results in \cite{BCJR:74,Li:95}. In this proof, we denote two events $\{{\bf s}[n-1]={\bf s}'\}$ and $\{{\bf s}[n]={\bf s}\}$ as ${E}_{n-1}({\bf s}')$ and ${E}_{n}({\bf s})$, respectively, for notational convenience. From \eqref{eq:def_state_set} and \eqref{eq:def_state_trans}, the marginal probability $\mathbb{P}({\bf x}[n]={\bf x}_k,\tilde{\bf Y})$ is rewritten as 

\vspace{-5mm}{\small{\begin{align}\label{eq:Adpx1:marginal}
	\mathbb{P}({\bf x}[n]={\bf x}_k,\tilde{\bf Y})
	&\overset{(a)}{=}\sum_{{\bf s}'\in \mathcal{S}_{n-1},{\bf s}\in\mathcal{S}_n}\mathbb{P}({E}_{n-1}({\bf s}'),{E}_{n}({\bf s}),{\bf x}[n]={\bf x}_k,\tilde{\bf Y}) \nonumber \\
	&\overset{(b)}{=}\sum_{({\bf s}',{\bf s})\in\mathcal{V}_n(k)}\mathbb{P}({E}_{n-1}({\bf s}'),{E}_{n}({\bf s}),\tilde{\bf Y}),
\end{align}}}for $k\in\mathcal{K}$, where the equalities in (a) and (b) are obtained from \eqref{eq:def_state_set} and \eqref{eq:def_state_trans}, respectively. Then by Bayes’ rule and the conditional independence, a pair-wise probability in \eqref{eq:Adpx1:marginal} is factorized as

\vspace{-5mm}{\small{\begin{align}\label{eq:Adpx1:pariwise}
	\mathbb{P}({E}_{n-1}({\bf s}'),{E}_{n}({\bf s}),\tilde{\bf Y}) 
	&=\mathbb{P}\big({E}_{n-1}({\bf s}'),{E}_{n}({\bf s}),\tilde{\bf Y}_1^{n-1},{\bf y}[n],\tilde{\bf Y}_{n+1}^{N_{\rm d}+L_{\mathcal D}-1}\big) \nonumber\\
	&=\mathbb{P}\big(\tilde{\bf Y}_{n+1}^{N_{\rm d}+L_{\mathcal D}-1}\big|{E}_{n-1}({\bf s}'),{E}_{n}({\bf s}),\tilde{\bf Y}_1^{n-1},{\bf y}[n]\big)
		\nonumber \\
	&~~~\times \mathbb{P}\big({\bf y}[n],{E}_{n}({\bf s})\big|\tilde{\bf Y}_1^{n-1},{E}_{n-1}({\bf s}')\big) 
		\mathbb{P}\big(\tilde{\bf Y}_1^{n-1},{E}_{n-1}({\bf s}')\big) \nonumber\\
	&=\underbrace{\mathbb{P}\big(\tilde{\bf Y}_{n+1}^{N_{\rm d}+L_{\mathcal D}-1}\big|{E}_{n}({\bf s})\big)}_{=\beta_{n}({\bf s})}
	\underbrace{\mathbb{P}\big({\bf y}[n],{E}_{n}({\bf s})\big|{E}_{n-1}({\bf s}')\big)}_{=\gamma_{n}({\bf s}',{\bf s})}
	\underbrace{\mathbb{P}\big(\tilde{\bf Y}_1^{n-1},{E}_{n-1}({\bf s}')\big)}_{=\alpha_{n-1}({\bf s}')},
	%=&\alpha_{n-1}({\bf s}')\gamma_{n}({\bf s}',{\bf s})\beta_{n}({\bf s}),  \label{eq:factorization}
\end{align}}}where $\tilde{\bf Y}_{a}^{b}\!=\!\big({\bf y}[a],{\bf y}[a\!+\!1],\cdots,{\bf y}[b]\big)$ is a subsequence of $\tilde{\bf Y}$ for $a\leq b \leq {N}_{\rm d}+L_{\mathcal D}-1$. The first factor in \eqref{eq:Adpx1:pariwise} is computed in a \textit{backward} recursive manner as follows:

\vspace{-7mm}{\small{\begin{align}
	\beta_{n-1}({\bf s}')\!&=\!\mathbb{P}\big(\tilde{\bf Y}_{n}^{N_{\rm d}+L_{\mathcal D}-1}\big|{E}_{n-1}({\bf s}')\big)
	=\!\sum_{{\bf s}\in\mathcal{S}_{n}}\!\mathbb{P}\big(\tilde{\bf Y}_{n}^{N_{\rm d}+L_{\mathcal D}-1},{E}_{n}({\bf s}) \big| {E}_{n-1}({\bf s}')\big)\nonumber\\
	&=\!\sum_{{\bf s}\in\mathcal{S}_{n}}\!\mathbb{P}\big(\tilde{\bf Y}_{n+1}^{N_{\rm d}+L_{\mathcal D}-1}\big|{\bf y}[n],{E}_{n-1}({\bf s}'),{E}_{n}({\bf s})\big)   
		\mathbb{P}\big({\bf y}[n],{E}_{n}({\bf s})\big|{E}_{n-1}({\bf s}')\big)\nonumber\\
	&=\!\sum_{{\bf s}\in\mathcal{S}_{n}}\!\mathbb{P}\big(\tilde{\bf Y}_{n+1}^{N_{\rm d}+L_{\mathcal D}-1}\big|{E}_{n}({\bf s})\big)   
		\mathbb{P}\big({\bf y}[n],{E}_{n}({\bf s})\big|{E}_{n-1}({\bf s}')\big)\nonumber\\
	&=\!\sum_{{\bf s}\in\mathcal{S}_{n}}\! \beta_n({\bf s})\gamma_n({\bf s}',{\bf s}), \label{eq:Apdx1:beta}
\end{align}}}for ${\bf s}'\!\in\!\mathcal{S}_{n-1}$, where the initial value is given by $\beta_{N_{\rm d}+L_{\mathcal D}-1}({\bf s})=\mathbb{I}\{{\bf s}={\bf 0}_{(L_{\mathcal D}-1)N_{\rm tx}}\}$. Similar to the above, the third factor in \eqref{eq:Adpx1:pariwise} is computed in a \textit{forward} recursive manner as follows:

\vspace{-7mm}{\small{\begin{align}
	\alpha_{n}({\bf s})&=\mathbb{P}\big(\tilde{\bf Y}_1^n,{E}_{n}({\bf s})\big) 
	=\sum_{{\bf s}'\in\mathcal{S}_{n-1}} \mathbb{P}\big(\tilde{\bf Y}_1^n,{E}_{n-1}({\bf s}'),{E}_{n}({\bf s})\big)\nonumber\\
	&=\sum_{{\bf s}'\in\mathcal{S}_{n-1}} \mathbb{P}\big({\bf y}[n],{E}_{n}({\bf s})\big|\tilde{\bf Y}_1^{n-1},{E}_{n-1}({\bf s}')\big)
		\mathbb{P}\big(\tilde{\bf Y}_1^{n-1},{E}_{n-1}({\bf s}')\big)\nonumber\\
	&=\sum_{{\bf s}'\in\mathcal{S}_{n-1}} \mathbb{P}\big({\bf y}[n],{E}_{n}({\bf s})\big| {E}_{n-1}({\bf s}')\big)\mathbb{P}\big(\tilde{\bf Y}_1^{n-1},{E}_{n-1}({\bf s}')\big)\nonumber\\
	&=\sum_{{\bf s}'\in\mathcal{S}_{n-1}} \gamma_n({\bf s}',{\bf s})\alpha_{n-1}({\bf s}') \label{eq:Apdx1:alpha},
\end{align}}}for ${\bf s}\!\in\!\mathcal{S}_n$, where the initial value is given by $\alpha_0({\bf s})= \mathbb{I}\big\{{\bf s}={\bf 0}_{(L_{\mathcal D}-1)N_{\rm tx}}\big\}$. Lastly, the second factor in \eqref{eq:Adpx1:pariwise} is computed as

\vspace{-7mm}{\small{\begin{align}
	\gamma_{n}({\bf s}',{\bf s})\!=&\mathbb{P}\big({\bf y}[n],{E}_{n}({\bf s})\big|{E}_{n-1}({\bf s}')\big) 
	=\mathbb{P}\big({E}_{n}({\bf s})|{E}_{n-1}({\bf s}')\big)\mathbb{P}\big({\bf y}[n]\big|{E}_{n-1}({\bf s}'),{E}_{n}({\bf s})\big) \nonumber\\
	=&\mathbb{P}\big({\bf x}[n]\!=\!({\bf s})_{1:N_{\rm tx}}\big)\mathbb{P}\big({\bf y}[n]\big|{E}_{n-1}({\bf s}'),{E}_{n}({\bf s})\big) \nonumber \\
	=&\begin{cases}
		\frac{1}{2^M}\mathbb{P}\big({\bf y}[n]\big|{E}_{n-1}({\bf s}'),{E}_{n}({\bf s})\big), &{\rm for}~n\in\{1,\ldots,N_{\rm d}\}, \\
		\mathbb{P}\big({\bf y}[n]\big|{E}_{n-1}({\bf s}'),{E}_{n}({\bf s})\big), & {\rm for}~n\in\{N_{\rm d}+1,\ldots,{N}_{\rm d}+L_{\mathcal D}-1\},
	\end{cases} \label{eq:Apdx1:gamma}
%=&p(x[n]\!=\!s[1])\prod_{i=1}^{N_{\rm r}} p(y_i[n]|s', s,{\bf h_i}),
%=&p(x[n]\!=\!s[1])\prod_{i=1}^{N_{\rm r}} Q\left({\sf Re}\{y_i[n]\}\frac{{\sf Re}\left\{{\bf h}_i^{\top} \begin{bmatrix}s[1] \\s'\end{bmatrix}\right\}}{\sqrt{\sigma^2/2}}\right) \label{eq:gamma}\\
%&\times Q\left({\sf Im}\{y_i[n]\}\frac{{\sf Im}\left\{{\bf h}_i^{\top} \begin{bmatrix}s[1] \\s'\end{bmatrix}\right\}}{\sqrt{\sigma^2/2}}\right), \nonumber
\end{align}}}for $({\bf s}',{\bf s})\in\bigcup_{k\in\mathcal{K}}\mathcal{V}_{n,k}$, while $\gamma_{n}({\bf s}',{\bf s})\!=\!0$ for $({\bf s}',{\bf s})\notin\bigcup_{k\in\mathcal{K}}\mathcal{V}_{n,k}$. 
Note that $\gamma_{n}({\bf s}',{\bf s})\!=\!0$ when $({\bf s}',{\bf s})\notin\bigcup_{{\bf x}\in\mathcal{X}}\mathcal{V}_n({\bf x})$. From the results in \eqref{eq:Adpx1:marginal}--\eqref{eq:Apdx1:gamma}, we obtain the expressions in \eqref{eq:marginalProb}--\eqref{eq:gamma}.

\end{document}